\documentclass[a4paper,11pt]{article}
\pdfoutput=1 
\usepackage{jcappub} 

\usepackage[T1]{fontenc} 
\usepackage{verbatim}
\usepackage{csquotes}

\usepackage[style=nature,backend=biber, sorting = none, doi=false, maxbibnames=8, minbibnames=3]{biblatex}

\usepackage[dvipsnames]{xcolor}

\usepackage{ae,aecompl}

\usepackage[english]{babel}

\usepackage{xcolor}

\usepackage{newtxtext,newtxmath}
\usepackage[T1]{fontenc}
\usepackage{ae,aecompl}
\usepackage{enumerate}
\usepackage[switch, modulo]{lineno}
\usepackage[toc,page]{appendix}

\usepackage{amssymb}

\usepackage{newtxtext,newtxmath}

\usepackage[T1]{fontenc}

\DeclareRobustCommand{\VAN}[3]{#2}
\let\VANthebibliography\thebibliography
\def\thebibliography{\DeclareRobustCommand{\VAN}[3]{##3}\VANthebibliography}

\usepackage{graphicx}	
\usepackage{amsmath}	
\usepackage{float}

\usepackage{caption}
\usepackage{subcaption}
\usepackage{blindtext}

\title{A measurement of the scale of homogeneity in the Early Universe}

\author[a,b]{Benjamin Camacho-Quevedo}
\emailAdd{camacho@ice.csic.es} 
\author[a,b]{Enrique Gazta\~naga}
\emailAdd{gaztanaga@gmail.com} 

\affiliation[a]{Institute of Space Sciences (ICE, CSIC), 08193 Barcelona, Spain}
\affiliation[b]{Institut d\'~Estudis Espacials de Catalunya (IEEC), 08034 Barcelona, Spain}

\abstract{
We present the first measurement of the homogeneity index, $\mathcal{H}$, a fractal or Hausdorff dimension of the early Universe from the Planck CMB temperature variations $\delta T$ in the sky. This characterization  of the isotropy scale is model-free and purely geometrical,
independent of the amplitude of $\delta T$.
We find evidence of homogeneity ($\mathcal{H}=0$) for scales larger than
$\theta_{\mathcal{H}} = 65.9 \pm 9.2 \deg $
 on the CMB sky.  This finding is at odds with the $\Lambda$CDM prediction, which assumes a scale invariant infinite universe.
Such anomaly is consistent with the well known low quadrupule amplitude in the angular $\delta T$ spectrum, but quantified in a direct and model independent way. We estimate the significance  of our finding for $\mathcal{H}=0$ using a principal component analysis from the sampling variations of the observed sky. This analysis is validated with theoretical prediction of the covariance matrix  \textcolor{black}{ and simulations, booth base purely on data or in the $\Lambda$CDM prediction.}
Assuming translation invariance (and flat geometry) we can convert the isotropy scale $\theta_\mathcal{H}$ into a (comoving) homogeneity scale which is very close to the trapped surface generated by the observed cosmological constant $\Lambda$.}


\begin{document} 
\maketitle
\flushbottom



\section{Introduction}

During the 1970s, Mandelbrot \cite{Mandelbrot,Mandel} showed how the fractal or Hausdorff dimension (introduced in 1918 by mathematician Felix Hausdorff) is a rigorous way  to characterize geometrical irregularities and chaos in Nature. The Hausdorff dimension  estimate is independent of the amplitude of such irregularities and it only depends on how they scale with size. These ideas have been applied to galaxy surveys as a way to test if the metric of the Universe is geometrically homogeneous.  Here we apply this concept for the first time to the early Universe. Because the amplitude of temperature fluctuations $\delta T/T$ directly relates to the gravitational potential through the Sachs-Wolf effect \cite{Sachs-Wolf}, the   $\delta T/T$ amplitudes are quite small, which does not necessarily mean that the universe is homogeneous. We need to explore how fluctuations change with scale. Therefore here we propose to do this in a model-independent way using the Hausdorff dimension, which we will adapt to the CMB case.

The Standard Cosmological model, the $\Lambda$CDM, relies on  the cosmological principle (CP): the Universe must be (statistically) homogeneous and isotropic on the largest scales. This implies that the quantities that we measure  are rotation and translation invariant. For the flat case, the metric can be written as 
the Friedmann-Lemaitre-Robertson-Walker (FLRW) metric:
\begin{equation}
ds^2 = -d\tau^2 + a^2 (\tau) \left[ d\chi^2 + \chi^2 
d\Omega^2. \right]
\label{eq:frw}
\end{equation}
Primordial photons from the surface of last scattering (a spacelike hypersurface with $d\tau \simeq 0$ and $d\chi \simeq 0$) must then follow:
\begin{equation}
ds^2 = d\Omega^2 \equiv d\theta^2 + \cos^2(\theta) d\phi^2
\label{eq:frw2}
\end{equation}
in units where $a \chi=1$.
Does our observed universe actually follow such metric on large scales? We will show in \S\ref{sec:H} how we can test this assumption using the Hausdorff dimension.

Homogeneity and isotropy are simple and logically consistent, so  it was postulated without any direct observational evidence,
even before we knew that the Universe was bigger than the 
Milky Way. Given the large amount of data collected by cosmic maps recently, the homogeneity and isotropy of the Universe can and should be tested directly \cite{Ellis}. Several authors have provided such tests using the distribution of galaxy tracers, such as LRGs and QSOs (e.g. \cite{Hogg,scrimgeour2012wigglez,alonso2014measuring, alonso2015homogeneity,laurent201614, ntelis2017exploring, gonccalves2018cosmic, gonccalves2018measuring, HomogeneityQuasars} and references therein). 
The fractal dimension (also called correlation dimension) has been used to characterize the scale at which the homogeneity is reached. Most of these works quote scales between 100 and 200 Mpc/h. Nevertheless these numbers reflect more the current (sampling variance) accuracy of galaxy maps than a direct geometrical measurement of the metric. Some works implement the counts-in-sphere algorithms to get information on the 3D space, but in reality this is restricted to the light-cone. The other way to test the homogeneity, in a model-independent way, is by computing the fractal dimension on a 2D radial shell (or fixed redshift). Such a 2D measurement  only gives us information about isotropy and it requires combination with the Copernican principle (or translation invariance) or measurements at different redshifts to explore 3D homogeneity.

Statistical isotropy has also been tested in different ways in the CMB (e.g. 
\cite{P13isotropy,P15isotropy,2015PhRvD..92f3008A,schwarz2016cmb,2016JCAP...06..042M,P18isotropy}), but not in a geometrical way using the Hausdorff dimension. The fact that the amplitude of fluctuations is small and consistent with $\Lambda$CDM is usually taken as evidence for the FLRW metric. 
There are some previous analysis on determining the fractality of the isophotes on the CMB temperature on COBE, WMAP and Planck (e.g. \cite{pino1995evidence, kobayashi2011fractal,myllari2016fractality}), but isophotes do not directly relate to the homogeneity scale or the metric.
Here, we look instead at the variation of temperature fluctuations with scale and we follow the same methodology employed to study the homogeneity of the matter distribution as traced by galaxies  (for an extensive review \cite{baryshev2005fractal}). 

 \textcolor{black}{Updating and extending earlier results, \cite{Copi2009}  confirmed the lack of correlations $w(\theta)$ in the CMB on scales larger than $\theta \simeq 60$ degrees, which seems at odds with the $\Lambda$CDM expectations. However \cite{Efstathiou2010} argue that such measurements can not be used to exclude the $\Lambda$CDM model. The issue is that the $\Lambda$CDM predictions have more power on the low $C_L$ power spectrum multipoles $L$ than the actual CMB measurements, this results in very large sampling variance error predictions for $w(\theta)$.  Thus, the significance of the discrepancy  depends on the model assumptions \cite{Gaztanaga2003}. Here we want to revisit this issue using the  Hausdorff dimension $\mathcal{H}$, which contrary to $w(\theta)$ or $C_L$ is independent of the amplitude of CMB temperature fluctuations. We will present a new covariance error estimate  which is model independent and validate it with theory and simulations.}

The paper will be structured as follows: in \S\ref{sec:mothodology} we  summarize and develop the quantities that we want to constrain. In  \S\ref{sec:measurements} will will describe briefly the data 
and explain all the details related with our measurements. In \S\ref{sec:results} we will summarize our main result to end in \S\ref{sec:Conclusions}
with conclusions and discussion.


\section{Method}
\label{sec:mothodology}

In this section we describe the path that we have followed to construct our main observable. At the same time we provide a review of the standard method used to measure the fractal dimension with galaxy surveys. It is important to notice that even when isotropy does not imply 3D homogeneity when it is measured only in a 2D subspace, it is equivalent to homogeneity when it is combined with the copernican principle (translation invariance). It also has the advantage of being a model-independent  estimation of homogeneity \cite{alonso2014measuring, alonso2015homogeneity, gonccalves2018cosmic}. 

\subsection{The 3D galaxy case}
\label{subsec:standar3D}
Following \cite{laurent201614}, the fractal correlation dimension can be determined by taking
\begin{equation}
    D_2(r) \equiv \frac{d\ln N(<r)}{d\ln r},
\label{Eq:fractal-dimension}
\end{equation}
where $N(<r)$ is the counts-in-sphere, which gives the average number of objects on a sphere of radius $r$ around a given object. Given an homogeneous distribution, we would have $N(<r)\propto r^3$, which makes $D_2(r)=3$ (for a space with three dimensions).\\

As the geometrical sampling of the survey is not perfect, \cite{scrimgeour2012wigglez} proposed  to correct this estimate by the completeness of the survey. The scaled or normalized counts-in-sphere $\mathcal{N}(<r)$  is introduced, which is just $N(<r)$ normalized by the same quantity in a random homogeneous distribution that takes into account selection effects. Then, we have that $\mathcal{N}(<r)\propto r^{D_2-3}$, which leaves
\begin{equation}
    D_2(r) = \frac{d\ln \mathcal{N}(<r)}{d\ln r} + 3.
    \label{Eq:fractal-dim-normalized}
\end{equation}
Where the 3 is added due to the fact that when the distribution of points is normalized, the correlation dimension is reduced by the dimensionality of the space in which they are embedded.\\

Now, instead of following \cite{laurent201614},  \cite{ntelis2017exploring} propose (in their Appendix A) a different estimator for the fractal dimension, which is shown to be less biased. As well as the one derived on \cite{laurent201614}, it makes use of the two-point correlation function (2PCF). The methodology on this case begins by taking the probability of finding a particle around another particle on the volume $dV$, which in fact depends on the 2PCF
\begin{equation}
    dN = \Bar{\rho}[1 + \xi(r)]dV,
\end{equation}
where $\Bar{\rho}$ is the mean density of particles and $\xi(r)$ the two-point correlation function. To obtain the counts-in-sphere, we needed to integrate
\begin{equation}
\begin{aligned}
    N(<r) &= \int dN = \int\Bar{\rho}\left[1+\xi\left(\overrightarrow{r'}\right)\right]dV \\
    &= 4\pi\Bar{\rho} \int_0^r[1+\xi(r')]r'^2dr',
    \label{Eq:N-int}
\end{aligned}
\end{equation}
where we have taken that $\xi(\overrightarrow{r}) = \xi(r)$. Assuming that for a random homogeneous distribution $\tilde{N}(<r)=\frac{4}{3}\pi r^3\Bar{\rho}$, we have
\begin{equation}
\begin{aligned}
    \mathcal{N}(<r) &= \frac{N(<r)}{\tilde{N}(<r)} = \frac{3}{r^3}\int_0^r[1+\xi(r')]r'^2dr'\\
    &= 1 + \frac{3}{r^3}\int_0^r\xi(r')r'^2dr',
\end{aligned}
\label{Eq:N-corr}
\end{equation}

\subsection{Homogeneity vs fractal index}
\label{sec:H}

Usually the fractal dimension is defined for distributions in n-dimensional Euclidean space:
\begin{equation}
ds^2 = dx_1^2 + dx_2^2 + \ldots + dx_n^2 
\end{equation}
where the volume element is $dV_n= dx_1 dx_2 \ldots dx_n$ which gives a volume $V_n = r^n$ for a n-dimensional cell of equal side $dx_n = r$.
For any metric we can generalize the fractal dimension to:
\begin{equation}
D_n(r) \equiv n \frac{d\ln{N(<r)}}{d\ln{V_n}} 
\label{eq:D_n}
\end{equation}
for euclidian metric: 
\begin{equation}
D_n(r) =\frac{d\ln{N(<r)}}{d\ln{r}}
\label{eq:eD_n}
\end{equation}
This works for non euclidean metrics like $S^n$, an hypersphere embedded in $R^{n+1}$. For $n=2$ it corresponds to the CMB sphere in Eq.\ref{eq:frw2}:
\begin{equation}
ds^2 = dx_1^2 + \cos(x_1)^2 dx_2^2 = d\theta^2  +  \cos(\theta)^2 d\phi^2 
\end{equation}
which has a volume element $dV_2=d\Omega = \cos(\theta) d\theta d\phi$, so that $V_2 = \Omega= 2\pi[1-\cos(\theta)]$ and then Eq.\ref{eq:D_n} in this case gives:
\begin{equation}
D_2 = 2 \frac{d\ln{N(<\theta)}}{d\ln{\Omega}} 
\end{equation}
As we mentioned in previous subsections, it is more convenient to work with normalized density ${\cal{N}}=N/\bar{N}$ which subtracts the volume from $N$. In such case, we measure the homogeneity index:
\begin{equation}
\mathcal{H} \equiv n\frac{d\ln{{\cal{N}}(<r)}}{d\ln{V_n}} = D_n - n
\label{eq:H}
\end{equation}
so that $\mathcal{H}=0$ for a homogeneous space with $D_n=n$ dimensions.  \textcolor{black}{ We will define the scale of homogeneity by the scale where we reach $\mathcal{H} =0$.}

\subsection{The CMB case}

We can adapt the same machinery used above to determine the scale of homogeneity in the galaxy distribution to the case of fluctuations in the temperature $T$.   \textcolor{black}{The CMB temperature  maps are given in terms of
$\Delta_T  \equiv \langle  T - \Bar{T} \rangle_{\text{pixel}}$  averaged in healpixels \footnote{\hyperlink{http://healpix.sourceforge.net} {http://healpix.sourceforge.net}} (for a HEALPix pixel definition see \cite{Healpix.2005ApJ...622..759G, Healpy.Zonca2019}). This can also be converted
to $\eta_T=\langle T \rangle_{\text{pixel}}/\Bar{T}$ using the known mean CMB $\Bar{T} \simeq 2.725$K (but note that results in this paper do not depend on the value of $\Bar{T}$).}
So we are not measuring the galaxy particle  over-density $\cal{N}$, but instead the  mean $T$ fluctuation as normalised counts-in-cells  $\cal{P}$ (defined in Eq.\ref{Eq:N-corr-w}). The temperature fluctuation correlation $W_\eta$ is:
\begin{equation}
\begin{aligned}
W_\eta (\theta) &\equiv  \langle\eta_T(\vec{\theta_1}) \eta_T(\vec{\theta_2})\rangle_{\text{sky}} \\
& = 
1 +   \langle\delta T(\vec{\theta_1}) \delta T(\vec{\theta_2})\rangle_{\text{sky}}  = 1 +  w(\theta)
    \label{Eq:w-t}
\end{aligned}
\end{equation}
where $\theta=|\vec{\theta_2}-\vec{\theta_1}|$ and $\delta T(\vec{\theta}) \equiv \eta_T(\vec{\theta})  -1= \Delta_T/\Bar{T}$.
We calculate $W_\eta(\theta)$ by counting pairs of pixels in the healpix maps weighted by $\eta_T$.
We define the Homogeneity index as in Eq.\ref{eq:H}:
\begin{equation}
    \mathcal{H}(\theta) = 2\frac{d\ln{\mathcal{P}(<\theta)}}{d\ln{\Omega(<\theta)}} = D_2 - 2
    \label{Eq:homogeneity-index-T}
\end{equation}
(see also \cite{alonso2014measuring} for a similar definition for galaxies).
Following the same methodology as in Eq.\ref{Eq:N-int}:
\begin{equation}
\begin{aligned}
    P(<\theta) &= \int dP = \Bar{T} \int  W_\eta\left(\phi\right) d\Omega \\
&    = 2\pi\Bar{T} \int_0^{\theta}   W_\eta\left(\phi\right)  \sin{\phi}d\phi,
    \label{Eq:N-int-w}
\end{aligned}
\end{equation}
where on the third equality we have just integrated the azimuthal coordinate. For a homogeneous distribution $\tilde{P}(<\theta)=2\pi\Bar{T}(1-\cos{\theta})$, so  we define $\mathcal{P}$ as:
\begin{equation}
    \mathcal{P}(<\theta) = \frac{P<\theta)}{\tilde{P}(<\theta)} = \frac{1}{1-\cos{\theta}} \int_0^{\theta} W_\eta(\phi) \sin{\phi}d\phi 
\label{Eq:N-corr-w}
\end{equation}
Eq.\ref{Eq:N-corr-w} is the 2D equivalent of   Eq.\ref{Eq:N-corr}.
Note how the homogeneity index $\mathcal{H}$ in Eq.\ref{Eq:homogeneity-index-T} 
is independent of the amplitude of the signal $W_\eta$ and it just reflects the geometrical scaling of volume as traced by temperature fluctuations.

\begin{figure}
\begin{center}
     \begin{subfigure}[b]{0.675\textwidth}
         \centering
         \includegraphics[width=\textwidth]{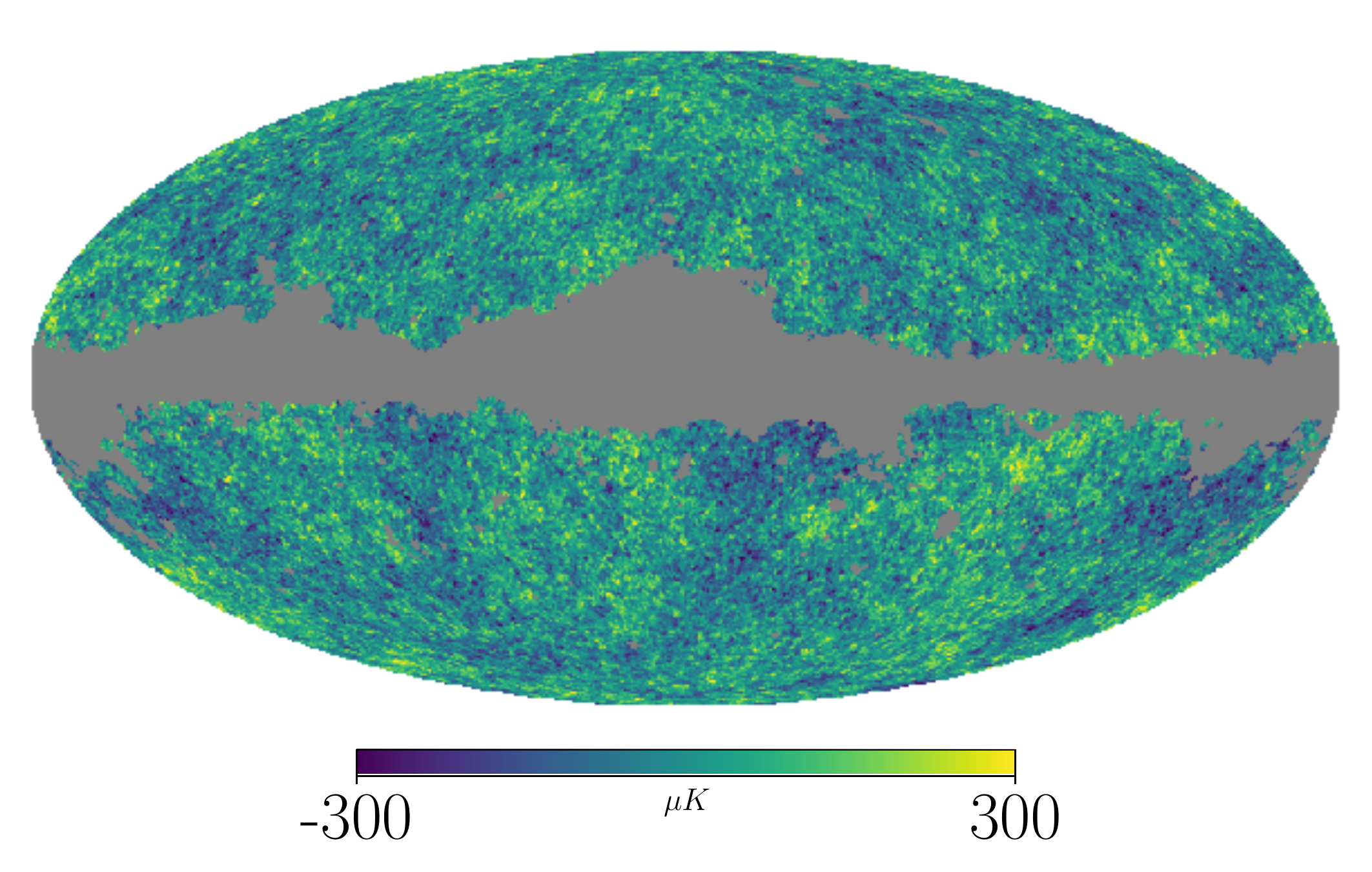}
         \caption{Observed Planck SMICA map.}
         \label{fig:smica-map}
     \end{subfigure}
     \begin{subfigure}[b]{0.675\textwidth}
        \centering
         \includegraphics[width=\textwidth]{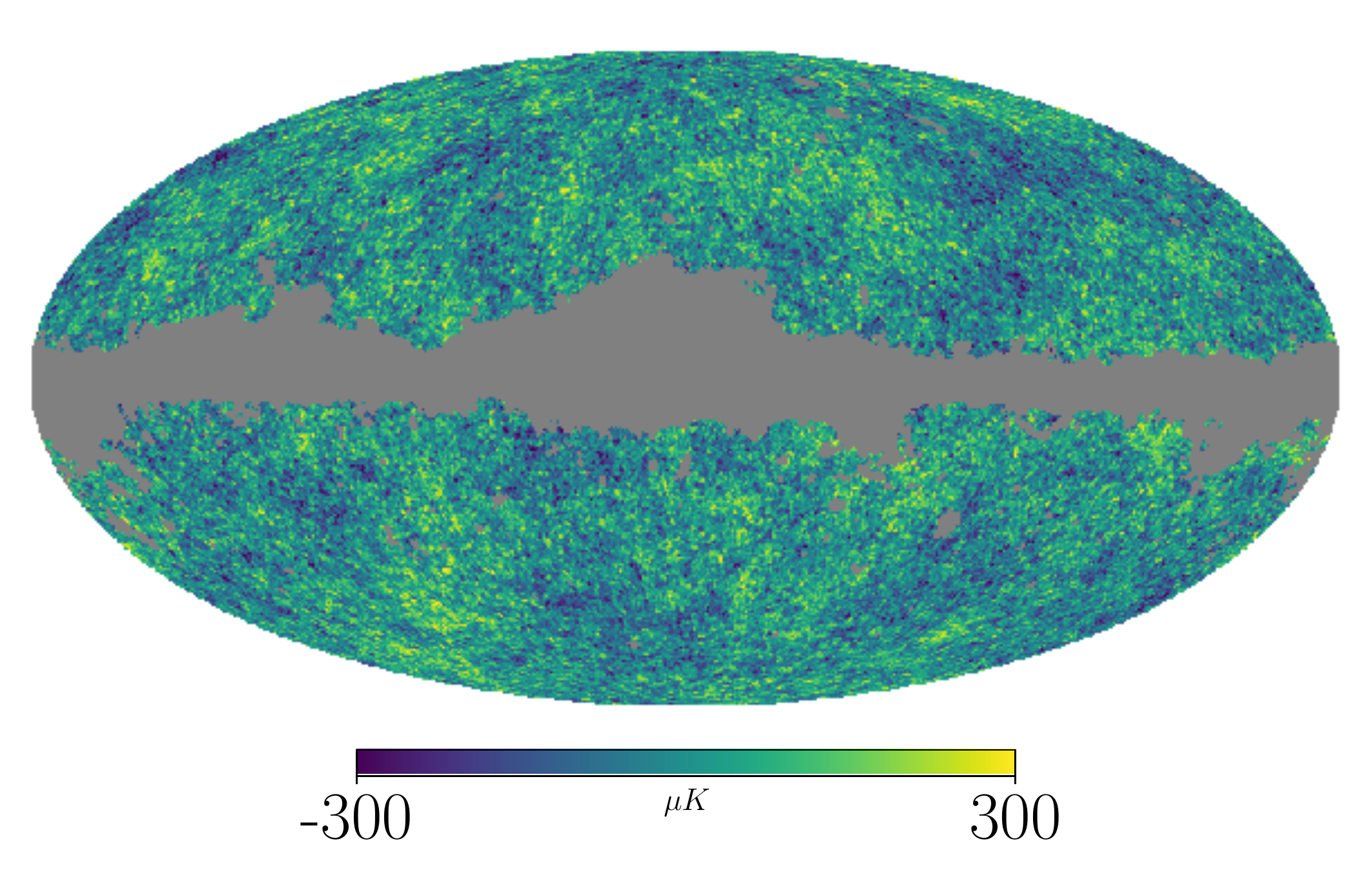}
         \caption{Obs. $C_\ell$ Sim. realization \#1.}
         \label{fig:cl-realization}
     \end{subfigure}
     \begin{subfigure}[b]{0.675\textwidth}
        \centering
         \includegraphics[width=\textwidth]{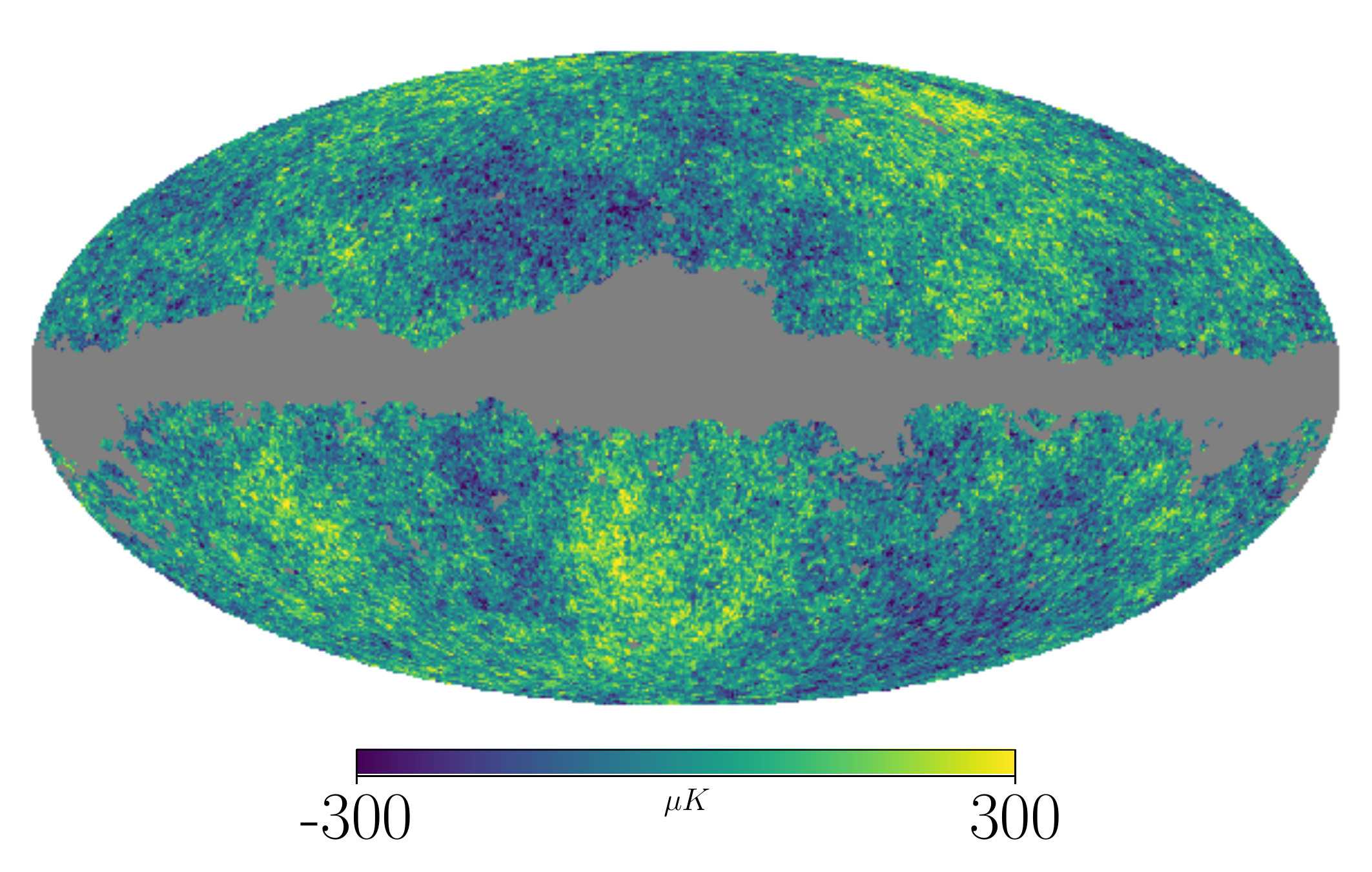}
         \caption{$\Lambda$CDM Sim. realization \#2.}
         \label{fig:lambda-realization}
     \end{subfigure}
\end{center}     
    \caption{CMB temperature maps from data, and simulations from the $C_\ell$ measured by Planck and $\Lambda$CDM best fit.}
    \label{fig:CMB-maps}
\end{figure}

\section{Data \& Measurements}
\label{sec:measurements}
We focus our analysis on the SMICA temperature maps of Planck-2018 \cite{akrami2018planck, akrami2020planck}, with its recommended mask. 
We use a HEALPix resolutions of $N{side}=128$ because we are interested mainly on the largest scales.
We have also used other maps, but find that the results are very similar, as can be seen in Table \ref{tab:PCA}. This is consistent 
with previous work, like \cite{schwarz2016cmb}, which showed that the $w(\theta)$ measurements, which are equivalent to $W_\eta(\theta)=1+w(\theta)$ here, are very robust against different choices. The results for $w(\theta)$ at large angular scales are strikingly similar to the COBE (\cite{COBEw2}) and WMAP results (\cite{Gaztanaga2003}), despite the enormous differences in time, technology, frequencies, and observational and analysis strategy. 

We followed these steps 
to estimate the homogeneity index:

\begin{itemize}
    \item \textbf{Reducing the Resolution:} The Temperature SMICA maps given by Planck are in a resolution of $N_{side}=2048$, then, we apply a downgrading on the resolution, rejecting pixels which contains a bad sub-pixel, which leaves a resolution of $N_{side}=128$. We apply the same procedure to the mask.
    
    \item \textbf{From pixels to sky coordinates:} We convert each pixel to its corresponding sky coordinates in the galactic system.
    
    \item \textbf{Jackknife resampling:} 
    We used the jackknife (JK) method to estimate errorbars from the same observed maps (\cite{Cabre07}).
    We are implementing the  \hyperlink{https://github.com/esheldon/kmeans_radec}{kmeans\_radec} (for an implementation example see \cite{kwan2016cosmology}) package on Python to divide the sky on 64 regions, with the same  approximate area, applying the $K$-means algorithm on the unit sphere.
    
    \item \textbf{Assigning temperature on the pixels as a weight:} Each one of the pixels is treated as an object with a weight that corresponds to the temperature fluctuation $\eta_T(\theta)$ on it, which allows us to compute $W_\eta= \langle\eta_T(\theta_i)\eta_T(\theta_j)\rangle$ by using weighted paircounts. 
    
    \item \textbf{Compute the paircounts using the public Python package \hyperlink{https://corrfunc.readthedocs.io/en/master/} {Corrfunc}:} Each pixel is a particle in {\it Corrfunc}\footnote{\hyperlink{https://corrfunc.readthedocs.io/en/master/}{ https://corrfunc.readthedocs.io/en/master/}} (\cite{Corrfunc.2020MNRAS.491.3022S}) with a weight $\eta_T$. Paircounts are determined between each JK region to implement the JK estimation. To study the numerical accuracy of our derivaties we have implemented two approaches, one with linear $\theta$-bins and the other in a logarithmic bins. 
    
    \item \textbf{Estimate the mean $W_\eta(\theta)$ over all JK regions:} We remove the paircounts contained in the $i$ JK region to estimate  $W{\eta i}(\theta)$. We then take the mean of such measurements and use individual   $W{\eta i}(\theta)$ to estimate the covariance matrix.
    
    \item \textbf{Integrate to obtain $\mathcal{P}(<\theta)$:} We are working with the mean number of paircounts estimated from the JK regions, then we put them on Eq.\ref{Eq:N-corr-w}, to compute the scaled integrated
    $\mathcal{P}(<\theta)$.
    
    \item \textbf{Differentiate to obtain $\mathcal{H}(<\theta)$:} Finally, we implement  Eq.\ref{Eq:homogeneity-index-T} as a numerical derivative. We have checked, using different bins and analytical models that our estimates are numerically stable. We find that we need to use a large number of angular bins to have a robust estimation of  $\mathcal{H}(<\theta)$. This has the disadvantage that different angular bins are strongly correlated, but we will account for that with the covariance matrix and the PCA and SVD analysis.
\end{itemize}

\subsection{CMB Maps and JK regions}
\begin{figure}
    \centering
    \includegraphics[width=14cm]{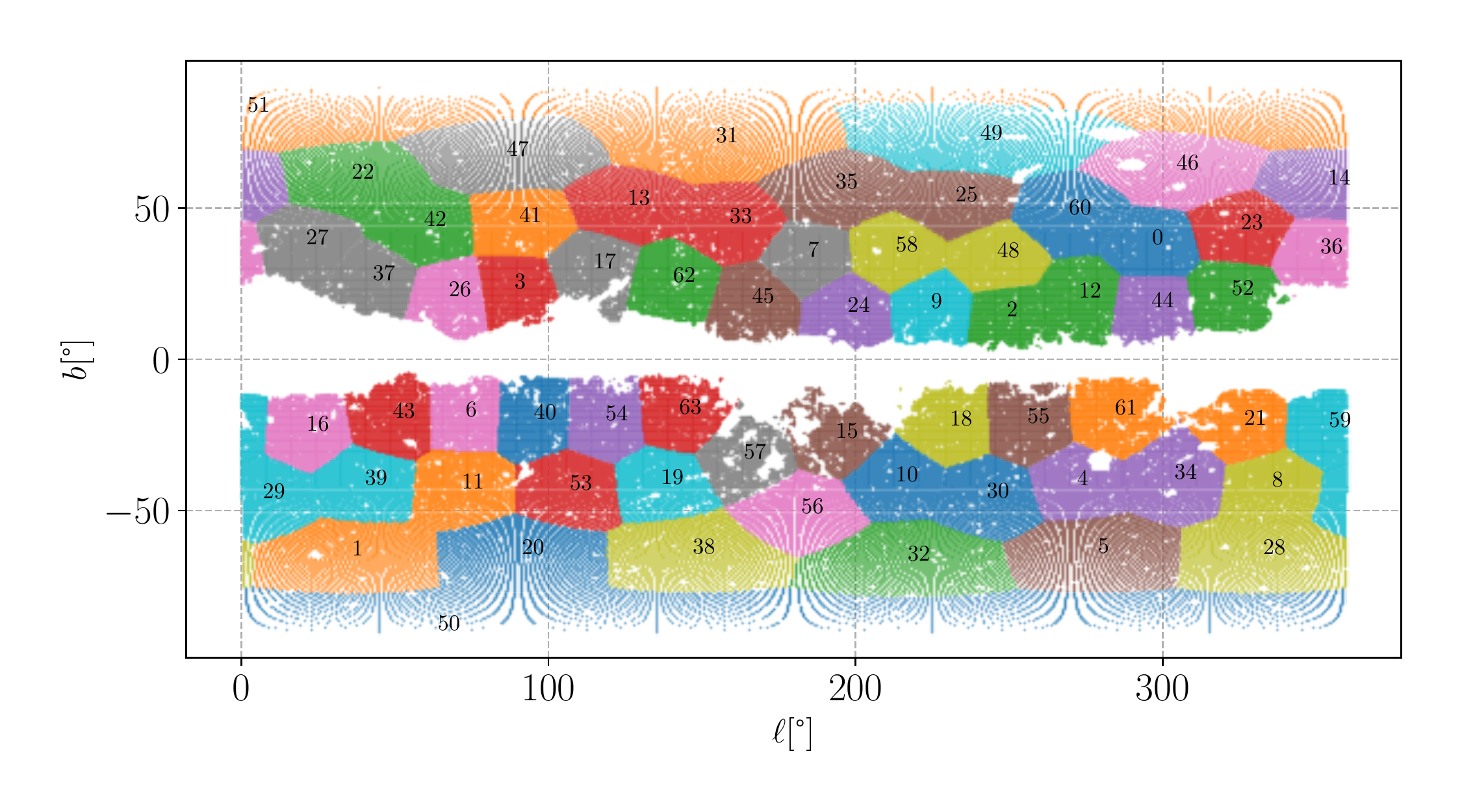}    
    \caption{Independent $N_{JK}=64$ JK regions in the sky.}
    \label{fig:JK-footprint}
\end{figure}
  \textcolor{black}{On Fig.\ref{fig:CMB-maps}, we show the SMICA map (\ref{fig:smica-map}), as well as a realization of the $C_\ell$ measured by Planck (\ref{fig:cl-realization}) and a $\Lambda$CDM one (\ref{fig:lambda-realization}).} Such maps are shown at a resolution $N_{side}=128$, we can see that even when the structure is blurred  on smaller scales, the large scale structure is preserved. Also, notice how contrary to the $\Lambda$CDM realization, a simulation from measured $C_\ell$  looks similar to the SMICA map.  Fig.\ref{fig:JK-footprint} illustrates how we divede the sky on 64 jackknife regions.
\begin{figure*}
    \centering
    \includegraphics[width=1.\textwidth]{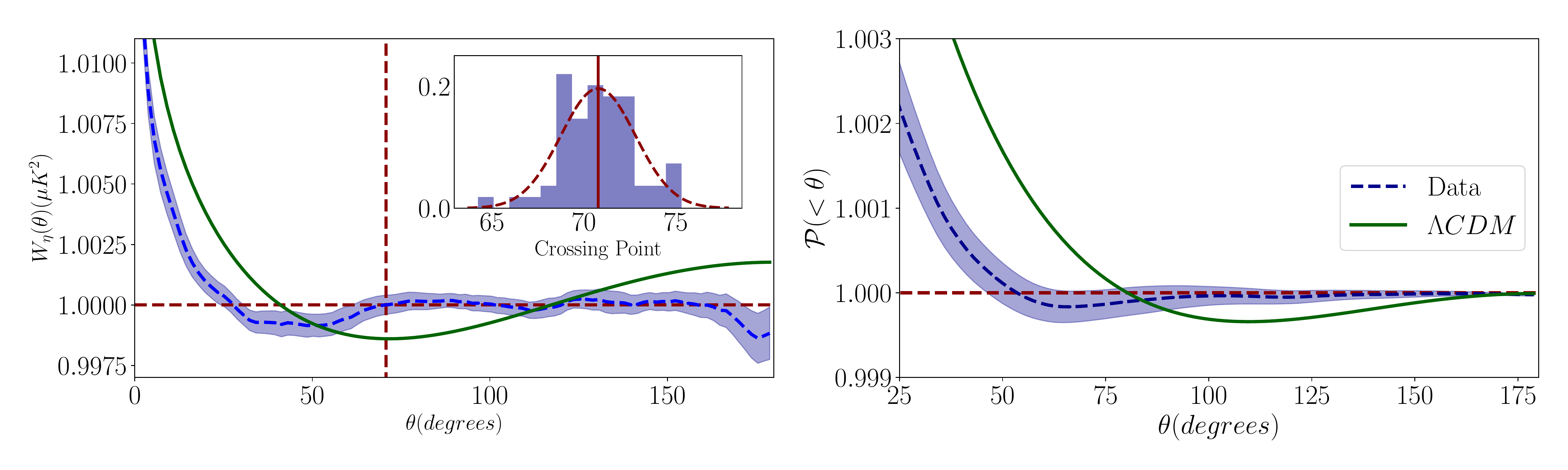}
\caption{The CMB two-point temperature correlation function  $W_\eta(\theta)$ in Eq.\ref{Eq:w-t}. The left panel compares measurements in Planck (dashed line) and sampling errors (blue shaded region) to the $\Lambda$CDM prediction (green continuous line). The right panel compares the $\mathcal{P}(\theta)$ (defined in Eq.\ref{Eq:N-corr-w}) CMB measurements in Planck (dashed line) and sampling errors (blue shaded region) to the $\Lambda$CDM prediction (green continuous line). On the homogeneous limit we should have $\mathcal{P}(\theta) = 1$.
The histogram on the left panel shows the distribution of the points at which $W_{\eta i}(\theta)$ crosses unity for a second time in each of the JK realizations (the vertical dashed red line indicates such point for the mean), where the mean of the distribution is indicated by the vertical line.}
\label{fig:w}
\end{figure*}
Fig.\ref{fig:w} shows the $W_\eta(\theta)= \langle\eta_T(\theta_i)\eta_T(\theta_j)\rangle$ measured on the CMB maps. 
The left panel shows (as dashed line) 
the mean over all the JK region. The mean and the error are estimated as:
\begin{equation}
    \Bar{W}_\eta(\theta) = \frac{1}{N_{JK}}\sum_i W_{\eta i}(\theta),
    \label{Eq:mean-w}
\end{equation}
and
\begin{equation}
\sigma^2= \frac{N_{JK}-1}{N_{JK}}\sum_i\left[ W_{\eta i}(\theta)-\Bar{W}_\eta(\theta) \right]^2 ,
\label{Eq:error-w}
\end{equation}
with $N_{JK}=64$, being the number of JK regions.  We have tried different values for $N_{JK}$ and find similar results (see below).
\cite{Cabre07} have shown that JK errors, the covariances (e.g. Eq.\ref{Eq:covariance}) and the SVD (e.g. Fig.\ref{fig:H-eigen}) of $w(\theta)$ agree well with  theoretical predictions and simulations for the same type of CMB healpix maps that we use here. This is a key ingredient in our estimation because we are able to calculate the homogeneity scale directly from data without any model assumption.  \textcolor{black}{Because the $\Lambda$CDM model has more power on large scales than real data, the $\Lambda$CDM model errors are larger (see \cite{Gaztanaga2003} or Fig.30 in \cite{Fosalba_2021}). But we are not trying to answer if observations could be a realization of 
the $\Lambda$CDM model, but rather what is the homogeneity scale derived just from observations without any model assumption.}  \textcolor{black}{Data covariances can be error-prone due to statistical fluctuations in the data, this is why we will compare to other estimations.}

The vertical line on the left panel on Fig.\ref{fig:w} shows the point at which $W_\eta$ crosses unity for a second time. This indicates where the correlations vanish, which is a well known anomaly on the CMB measurements \cite{Bennett-wmap,Gaztanaga2003,Copi2009,schwarz2016cmb}. The angle at which $W_\eta(\theta)$ flattens is $70.54 \deg$, and the mean JK value of this \textit{``crossing''} points is $70.77 \deg$, with a standard deviation of $16.14\deg$. The distribution of each one of this points per each JK is shown on the sub-panel of the left hand side of Fig.\ref{fig:w}. Recall that true scatter in Eq.\ref{Eq:error-w} is $\sqrt{N_{JK}-1}$ larger than the dispersion between the JK regions.

\begin{figure*}
         \centering
         \includegraphics[width=\textwidth]{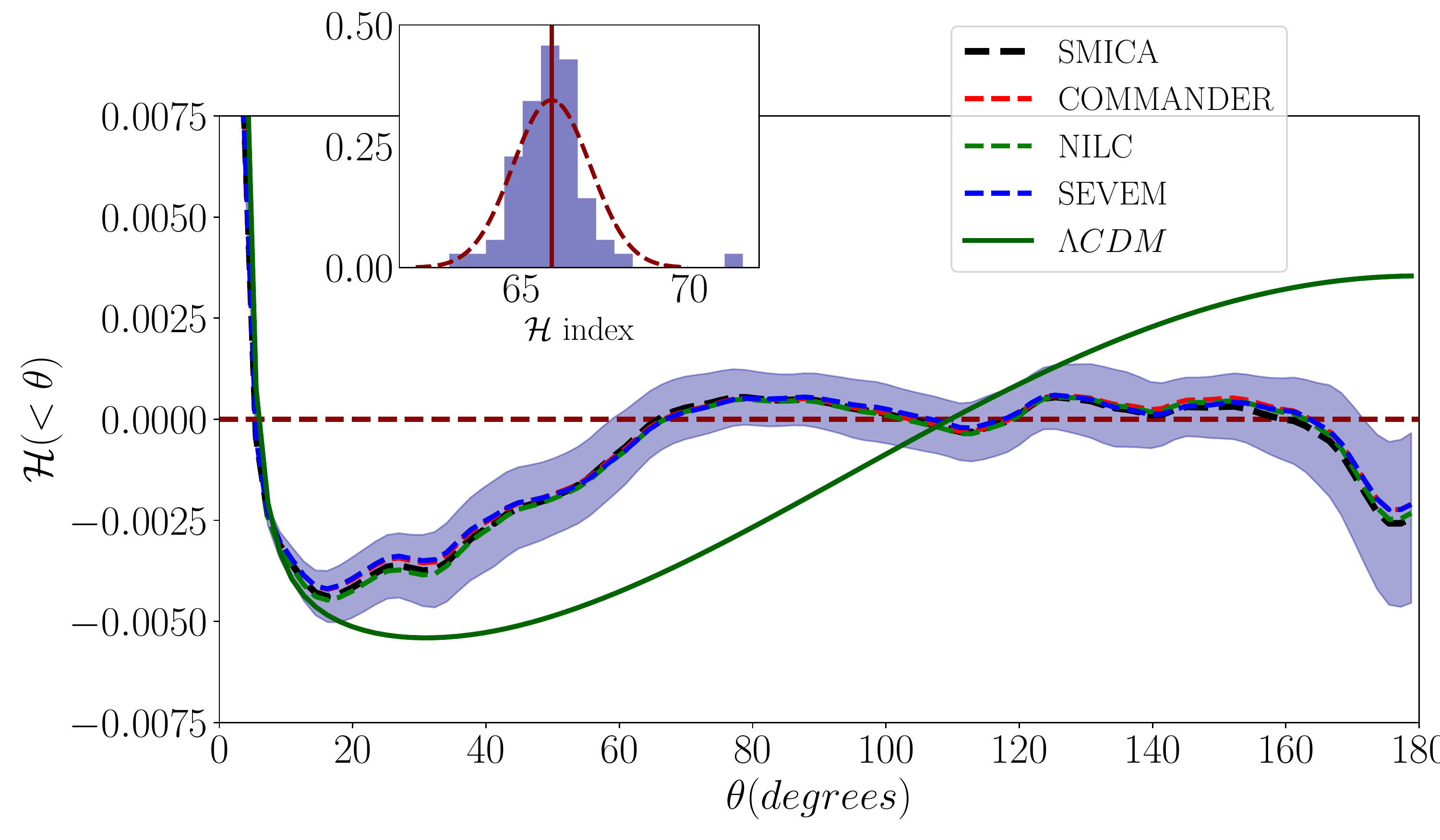}
         \caption{Homogeneity index $\mathcal{H}$ defined in Eq.\ref{Eq:homogeneity-index-T}. 
        Homogeneity is reached at $\mathcal{H}=0$, it is indicated by the horizontal red dashed line. The histogram on the sub-panel shows the distribution of the points at which $\mathcal{H}$ is flattened for each JK, where the mean is indicated by the vertical line. The different lines indicate the measurement obtained from different CMB maps, and the $\Lambda$CDM prediction, as indicated by the labels.} 
        
    \label{fig:H-linear}
\end{figure*}

After measuring $W_\eta$, the next step on our pipeline is to compute the normalised counts-in-cells from such measurements. We compute $\mathcal{P}(<\theta)$ using the estimator on Eq.\ref{Eq:N-corr-w}, which is shown on the right panel on the Fig.\ref{fig:w}. The Fig.\ref{fig:H-linear} shows the estimate of $\mathcal{H}(<\theta)$ in
Eq.\ref{Eq:homogeneity-index-T}. The red dashed horizontal lines show the expected value on the homogeneity limit for both quantities. On the $\mathcal{P}(<\theta)$ and $\mathcal{H}(<\theta)$ cases, the mean and error bars were estimated working with the equivalent expressions as for the equations \ref{Eq:mean-w} and \ref{Eq:error-w}.

\section{Results}
\label{sec:results}

\begin{figure*}
    \centering
     \begin{subfigure}{0.32\textwidth}
         \centering
         \includegraphics[width=\textwidth]{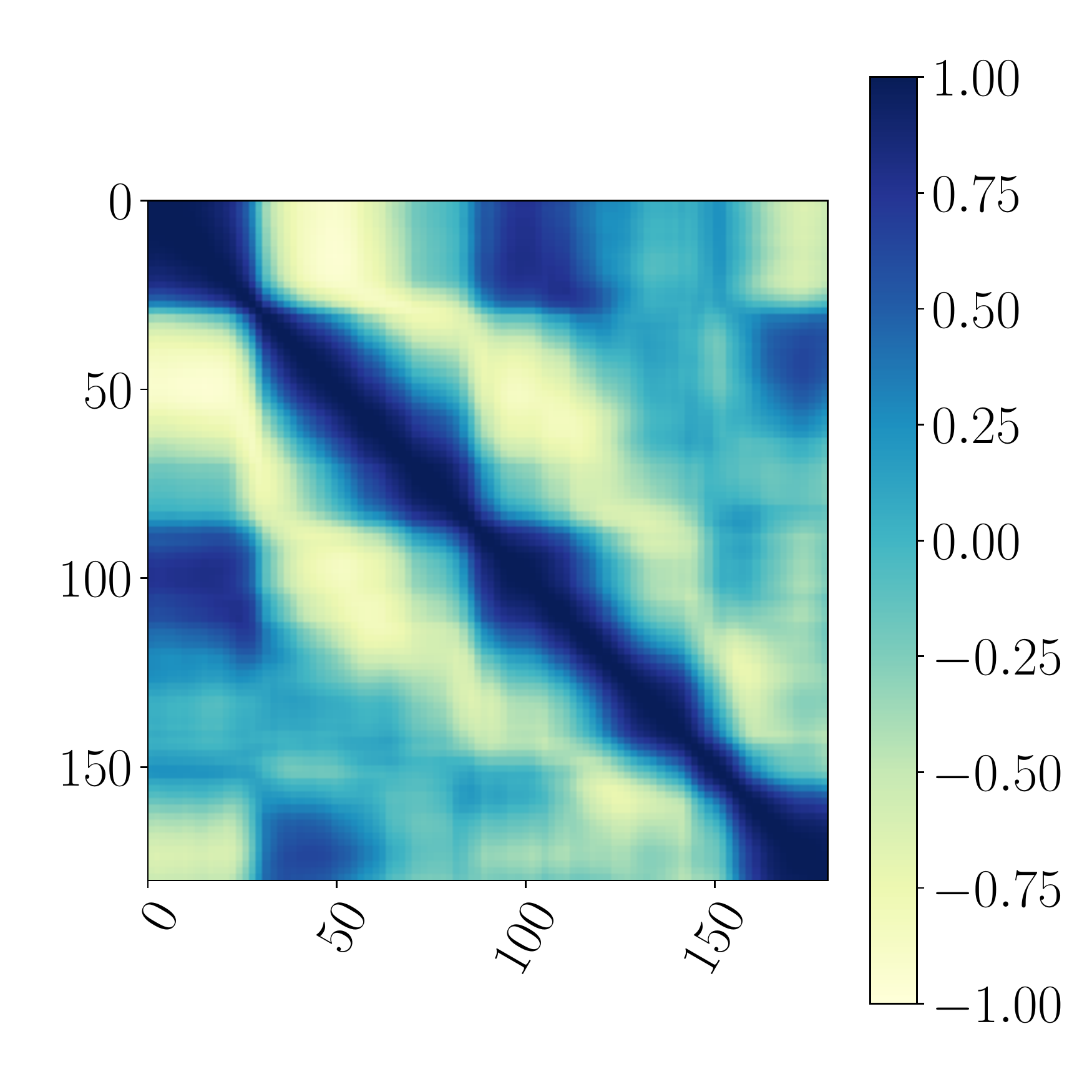}
         \caption{$\hat{C}_{ij}$ for $W_\eta(\theta)$.}
         \label{fig:w_cov}
     \end{subfigure}
     \begin{subfigure}{0.32\textwidth}
        \centering
         \includegraphics[width=\textwidth]{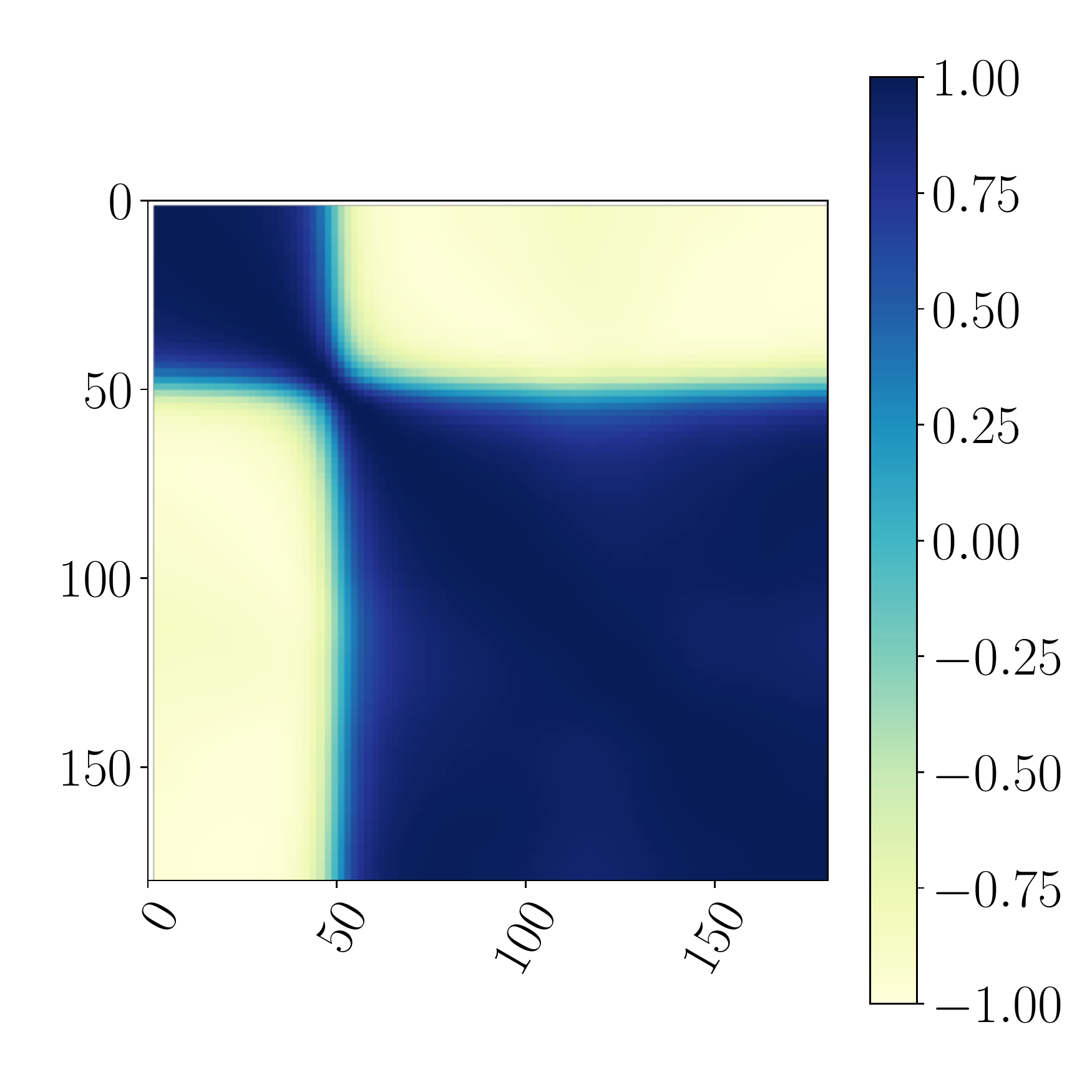}
         \caption{$\hat{C}_{ij}$ for $\mathcal{P}(\theta)$.}
         \label{fig:P_cov}
     \end{subfigure}
     \begin{subfigure}{0.32\textwidth}
        \centering
         \includegraphics[width=\textwidth]{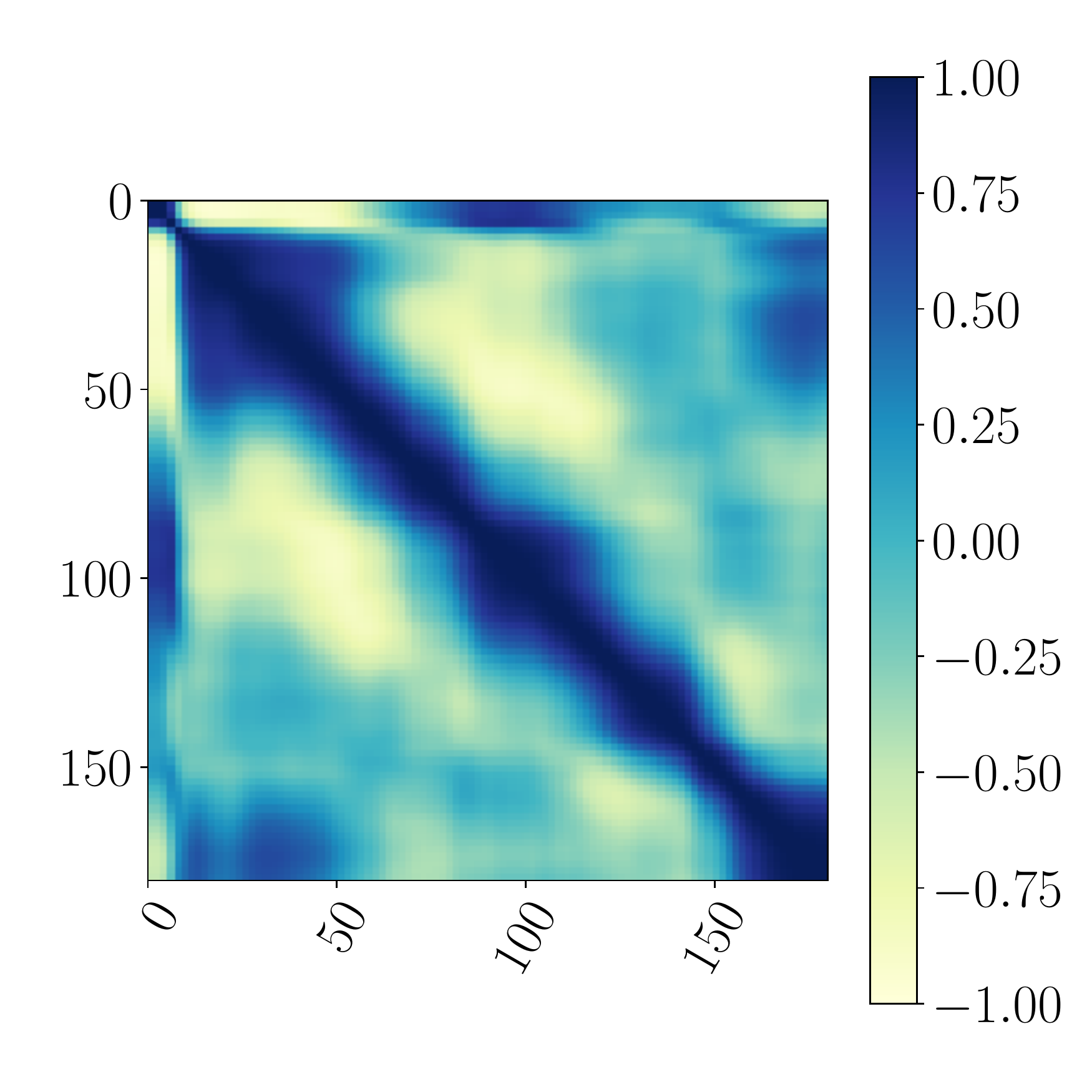}
         \caption{$\hat{C}_{ij}$ for $\mathcal{H}(\theta)$.}
         \label{fig:H_cov}
     \end{subfigure}
    \caption{Correlation matrices for $W_\eta(\theta)$, $\mathcal{P}(\theta)$ and $\mathcal{H}(\theta)$ estimated from the JK sampling.}
    \label{fig:covariances}
\end{figure*}

\subsection{Angular correlations}

As we have said before, there is a well known inconsistency between what $\Lambda$CDM predicts and what is measured on the 2PCF \cite{Gaztanaga2003, Bennett-wmap, schwarz2016cmb} related with the lack of correlation at scales larger than $60$ degrees. On Fig.\ref{fig:w}, the green line shows the temperature fluctuation correlation $W_\eta$ using the best fit $\Lambda$CDM spectrum given by Planck \cite{2020A&A...641A...5P} and using the relation

\begin{equation}
W_\eta(\theta) = 1 +   w(\theta)=1 + \sum_{\ell=1}\left( \frac{2\ell + 1}{4\pi}\right)P_{\ell}[\cos{(\theta)}]C_{\ell}
    \label{eq:transorm-cl2w}
\end{equation}
where $C_\ell$ is the angular power spectrum and $P_\ell$ the Legendre polynomials of degree $\ell$. 
The  $\Lambda$CDM  model is assumed to be scale invariant over an infinite volume and it therefore has structure on all CMB scales.
This is consistent with previous results and it is well discussed on \cite{schwarz2016cmb}. \\

\subsection{The homogeneity scale $\chi_{\mathcal{H}}$}

Although homogeneity is assumed in cosmology, it is not clear at which scale it should be reached. Here, as explained in \S\ref{sec:H}, we have taken the definition of homogeneity as the scale at which $\mathcal{H}$ is flattened, i.e. $\mathcal{H}=0$ is a stable value. 
This is illustrated in  Fig.\ref{fig:H-linear}.
In order to estimate the uncertainty in such scale, we have determined it on each one of the JK region measurements, whose distribution is shown on the histogram of the sub-panel, which gives us a scale of homogeneity of
\begin{equation}
    \theta_{\mathcal{H}} = \theta (\mathcal{H}=0) = 65.897 \pm 9.172 \deg 
\end{equation}
Note that these values are different from the ones we obtained from $W_\eta$
in Fig.\ref{fig:w}. In particular, the errors are smaller.
The scale $\theta_{\mathcal{H}}$ corresponds to a lower bound for the scale of homogeneity. Though this does not give a really precise estimation of the scale at which homogeneity is reached, it, in fact, tells us that departures from homogeneity from this scale and above are not seen. For this case, we see homogeneity at scales  $\theta_{\mathcal{H}}\gtrsim 60$. For $k=0$ and $\Omega_\Lambda =0.75$ this corresponds to a homogeneity distance $ \chi_{\mathcal{H}}$:
\begin{equation}
 \chi_{\mathcal{H}} \gtrsim \theta_{\mathcal{H}} d_A \simeq 3.3 c/H_0  ~~ ; ~~d_A = \int_{a_{\textrm{CMB}}}^1 \frac{da}{a^2 H(a)} \simeq 3.2 c/H_0
 \label{eq:dA}
\end{equation}
where $d_A$ is the comoving angular diameter distance to the CMB: $a_{\textrm{CMB}} \simeq 10^{-3}$ and $H(a) \equiv \dot{a}/a$ is the Hubble law.
In \S\ref{sec:comparison} we will give some interpretation for such scale..

The Fig.\ref{fig:H-linear} shows how, contrary to real data,  $\mathcal{H}$ for 
the  $\Lambda$CDM  model does not go to zero.  This is because geometrical homogeneity is never reached for  $\Lambda$CDM because
it assumes a much larger scale invariant (i.e. fractal) universe. Given that there is a strong covariance in these measurements, are these results significant?

\begin{figure*}
     \begin{subfigure}[b]{1.\textwidth}
         \centering
         \includegraphics[width=\textwidth]{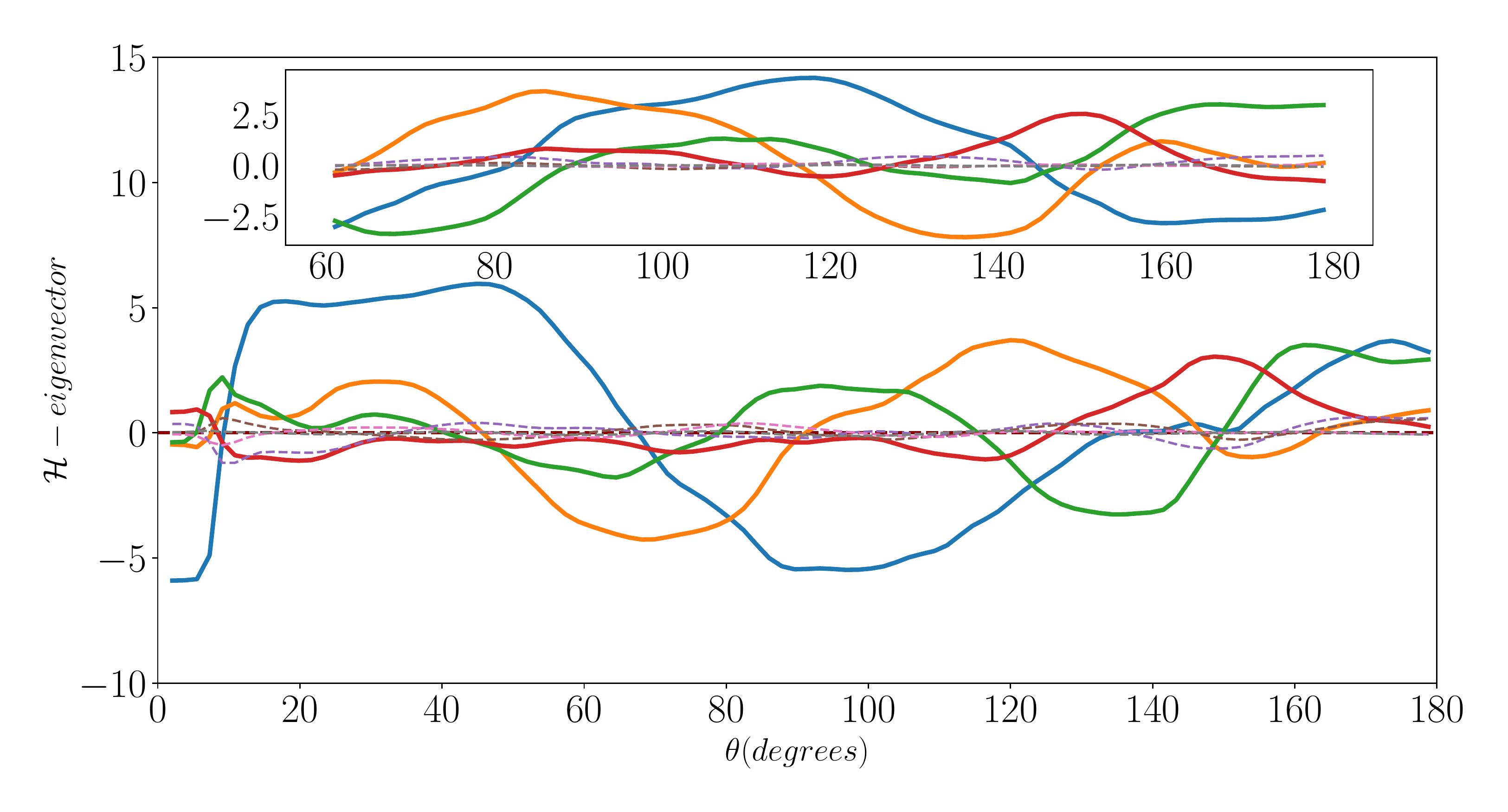}
         \caption{Weighted eigen-vectors.}
         \label{fig:H-eigen}
     \end{subfigure}
     \begin{subfigure}[b]{1.\textwidth}
        \centering
         \includegraphics[width=\textwidth]{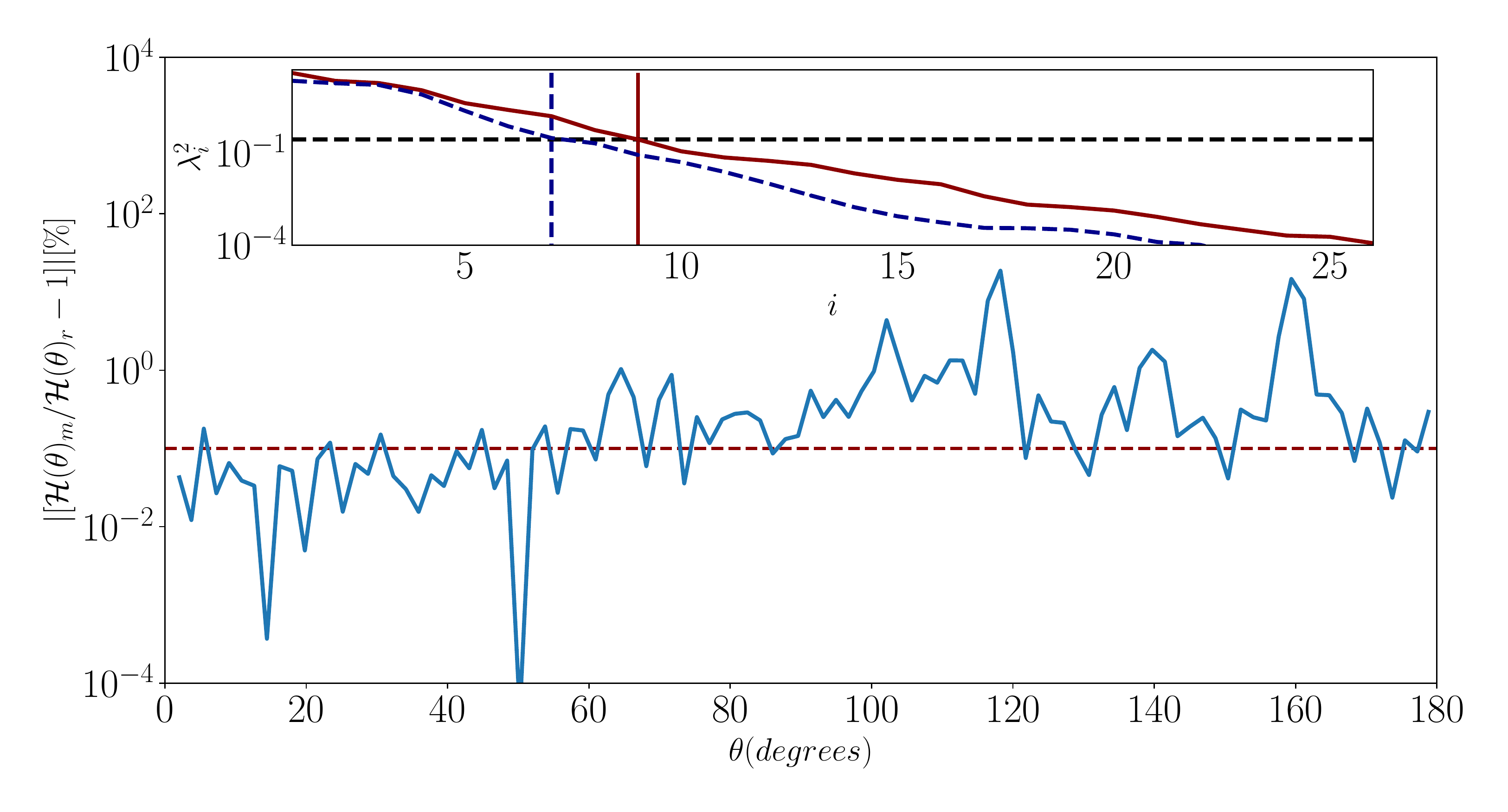}
         \caption{Comparison of the reconstructed and measured $\mathcal{H}(\theta)$.}
         \label{fig:H-rec}
     \end{subfigure}
\caption{Principal Components Analysis (PCA) using Singular Value Decomposition (SVD). On \ref{fig:H-eigen}, we show the dominant eigen-modes used to reconstruct $\mathcal{H}$ scaled by the amplitude of the corresponding eigenvalue.
The sub-panel shows the corresponding  eigen-modes when we restrict the analysis to $\theta \geq 59$. The continuous lines show the first four PC of our decomposition. The panel \ref{fig:H-rec} shows the relative difference between the measurement of $\mathcal{H}$ and the reconstruction using the dominant eigenvalues. The sub-panel shows the eigenvalues on decreasing order for all sky (continuous red) and for $\theta \geq 59$ (dashed blue), the vertical lines show numbers 7 and 9, which correspond to where we cut in the dominant values (indicated by the black dashed horizontal line) used in the SVD for each case. }
\label{fig:PCA-results}
\end{figure*}

\subsection{Significance and PCA}
\label{sucsec:pca}

In order to estimate the significance of our measurements compared with a set of different assumptions (model motivated) we have made a Principal Component Analysis (PCA). First, we have taken the covariance matrix for each one of the estimated parameters as
\begin{equation}
    C_{ij} = \frac{N_{JK}-1}{N_{JK}}\sum_{i,j} \left(X_i-\Bar{X_i}\right) \left(X_j-\Bar{X_j}\right),
    \label{Eq:covariance}
\end{equation}
with $\Bar{X}$ as the mean value of the measurement in a given angular bin over all the JK realizations. On the figure \ref{fig:covariances} we show the correlation matrices (or normalized covariance), $\hat{C}_{ij}=C_{ij}/\sqrt{C_{ii}C_{jj}}$,
estimated for the correlation function, the normalised count-in-cells and the homogeneity index. On the panel \ref{fig:P_cov}, a strong covariance can be seen  in $\mathcal{P}(<\theta)$, which is expected, since it is an integral over a wide range of $\theta$. Nevertheless, these covariances disappear for $\mathcal{H}(<\theta)$, which is shown in the panel \ref{fig:H_cov}, where the correlation matrix is similar to the one corresponding to $W_\eta(\theta)$ on the panel \ref{fig:w_cov}.

In an idealized scenario, we could compute the  $\chi^2$ for different models by solving 
\begin{equation}
    \chi^2=\sum_{i=1}^{i=N_b}\sum_{j=1}^{j=N_b}\Delta_i\hat{C}_{ij}^{-1}\Delta_j
    \label{eq:chi2}
\end{equation}
where $\Delta_i=\left(\mathcal{H}^{obs}(i)-\mathcal{H}^{mod}(i)\right)/\sigma_{\mathcal{H}}(i)$ and $\mathcal{H}^{mod}$ denotes the different implemented models. 
Due to the fact that $N_{JK}$ is small, this  gives raise to uncertainties in the covariance matrix: $\Delta C_{ij}\simeq \sqrt{\frac{2}{N_{JK}}}$  \cite{gaztanaga2005three}.
Because angular bins are strongly correlated there are degeneracies and we need to perform a PCA to be able to invert $\hat{C}_{ij}$.
To account for such artifacts we perform a PCA by taking the Singular Value Decomposition (SVD, for a more extended discussion, see \cite{gaztanaga2005three,Cabre07}):
\begin{equation}
    \hat{C}_{ij} = (U_{ik})^{\dagger}D_{kl}V_{lj}
    \label{eq:pca-decomposition}
\end{equation}
Where $D_{ij}=\lambda^2\delta_{ij}$ is the singular values diagonal matrix and $U_{ik}, V_{lj}$ are orthogonal matrices that contain the corresponding eigen-vectors, which besides being adequate to separate the signal from the noise, enable us to project these measurements on a diagonal subspace with a smaller dimensionality by taking
\begin{equation}
    \bar{\mathcal{H}}(i)=\sum_iU_{ij}\frac{\mathcal{H}(j)}{\sigma_{\mathcal{H}}(j)}.
    \label{eq:pca-projection}
\end{equation}
In this way, we can work just with the dominant modes on the new subspace such that they satisfy 

\begin{equation}
    \lambda^2 > \sqrt{\frac{2}{N_{JK}}}.
    \label{eq:eigenmodes}
\end{equation}
The $\mathcal{H}$-eigenvectors, computed with the Eq.\ref{eq:pca-projection} correspond to 
a new base where the covariance is diagonal. They are shown in Fig.\ref{fig:H-eigen}, where we have weighted them by their corresponding eigenvalue to illustrate which are the dominant contributions (we have highlighted the first 4 dominant eigen-vectors). We have performed two different SVD and PCA, for all scales and using only $\theta > 59$ degrees, where
$\mathcal{H}$ is close to zero.
In Fig.\ref{fig:H-rec}
we show that the reconstruction with a subset of the dominant components recovers well the original signal within $1\%$. In the sub-panel of this figure we also show the ordered amplitude of the eigenvalues, the vertical lines show the eigenvalue at which we stop to take into account their contribution in Eq.\ref{eq:eigenmodes}, and the horizontal dashed line its amplitude.

    
We use different models for $\mathcal{H}^{mod}$ to compare to data and validate our results. We estimated the $\chi^2$  in the new basis as:
\begin{equation}
\chi^2 = \sum_i^{\#\textrm{d.o.f.}} \hat{\Delta}_i^2/\lambda_i^2
\end{equation}
where $\hat{\Delta}_i \equiv  \sum_j U_{ij}\Delta_j$ is the difference between model and measurements projected into the SVD eigen-vectors. We tried several models for ${\mathcal{H}}^{mod}$:
\begin{enumerate}
    \item Homogeneous distribution for all angles, i.e. $\mathcal{H}^{mod}=0$.
    \item Homogeneous distribution  $\mathcal{H}^{mod}=0$ for $\theta \geq$ 59 degrees.
   \item $\Lambda$CDM for all angles.
    \item $\Lambda$CDM:   $\mathcal{H}^{mod}=\mathcal{H}_{\Lambda CDM}$ for $\theta \geq$ 59 degrees.
\end{enumerate}

On the cases when we used scales $\geq 59$ degrees, we performed the SVD taking $\hat{C_{ij}}$ only for such scales. The sub-panel on Fig.\ref{fig:H-eigen} shows the eigen-vectors obtained on this case and the blue dashed line in the sub-panel on Fig.\ref{fig:H-rec} the corresponding eigenvalues ordered, the vertical dashed line points the eigenvalue at which we stop to take into account their contribution.

The results are summarized in the Table \ref{tab:PCA}. We see that the $\Lambda$CDM prediction ( \textcolor{black}{last column}) 
is not consistent with our Planck measurements at scales $\geq 59$ degrees. On the other hand, the results using all scales are dominated by the small scale angles which have larger S/N. On that case, the data is in general consistent with the $\Lambda$CDM model, as expected.

Measurements discard homogeneity for all scales at $\sim 10 \sigma$  (per d.o.f.). However, there is  strong evidence for a model that assumes homogeneity at $\theta > 59$ degrees.

\begin{table}
	\centering
	\caption{This table summarizes the results of our tested models. The degrees of freedom ($\#d.o.f.$) are given by the number of  modes that we have used to invert $\hat{C}_{ij}$. The first column indicates the map that we have used to perform the homogeneity index $\mathcal{H}$ measurement; the second one indicates the method we have implemented to compute the covariance; and the remaining two columns indicate the $\chi^2/\#d.o.f.$ 
	  \textcolor{black}{and corresponding probabilies}
	for the two hypothesis  ($\mathcal{H}=0$ or $\mathcal{H}=\mathcal{H}_{\Lambda\text{CDM}}$) for scales $\theta >$ 59 degs. On the second column the $N_{JK}$ label indicates when the covariance was computed implementing JK and the number of regions that we have taken; Theory indicates that we have implemented the analitical expresions, using as input the $C_\ell$ directly measured by Planck (Obs.), or the one given by the best-fit model ($\Lambda$CDM); Sim. indicates when the covariance was estimated from simulations applying or not the observed mask to the  simulated maps.}
	\label{tab:PCA}
	\begin{tabular}{cccc} 

	        	     Mean & Error Covariance & $\mathcal{H}[\theta>59]=0$ &
	        	     $\mathcal{H}[\theta>59]=\mathcal{H}_{\Lambda\text{CDM}}$ \\
	    \hline
		\hline
(masked) & \bf{MODEL INDEPENDENT}  &  
  \multicolumn{2}{c}{$\chi^2\,/\,\# d.o.f.$ (Probability)}   \\
\hline
		SMICA & $N_{JK}=64$   &  8.7 / 7 (27\%)  & 61.4 / 7  ($<10^{-5}$) \\
		SMICA & $N_{JK}=128$   & 5.1 / 7 (64\%) & 81.3 / 7  ($<10^{-5}$) \\
		COMMAND & $N_{JK}=64$   & 8.0 / 7 (33\%)   & 57.9 / 7  ($<10^{-5}$) \\
		SEVEM & $N_{JK}=64$   & 6.5 / 7 (48\%)  & 197.5 / 7   ($<10^{-5}$)\\
		NILC  & $N_{JK}=64$  & 5.3 / 7 (62\%)   & 65.6 / 7  ($<10^{-5}$)\\
		SMICA & Theory (Obs.) no mask   &  6.1  / 4  (19\%) & 201.2 / 4  ($<10^{-5}$)  \\
		SMICA & Theory (Obs.) masked   & 4.8  / 4 (30\%)  & 156.7 / 4  ($<10^{-5}$) \\
		SMICA  & 128 Sim. no mask   & 4.9/ 4 (29\%)   & 161.7 / 4   ($<10^{-5}$)\\
		SMICA  & 128 Sim. masked    & 3.2 /  4 (52\%)  & 90.4 / 4   ($<10^{-5}$)\\
		SMICA  & 256 Sim. masked    & 3.2 /  4 (52\%)   & 107.8 / 4   ($<10^{-5}$)\\
		Sim. Mean  & 256 Sim. masked    & 4.4 /  4 (35\%)  & 87.3 / 4  ($<10^{-5}$) \\
		Sim. masked \#1  & 256 Sim. masked    & 0.4 /  4 (98\%)   & 116.7 / 4  ($<10^{-5}$) \\
		Sim. masked \#2  & 256 Sim. masked    & 4.7 /  4 (32\%)   & 117.3 / 4  ($<10^{-5}$) \\
		Sim. masked \#3  & 256 Sim. masked    & 1.8 /  4 (77\%)   & 105.6 / 4  ($<10^{-5}$) \\		
	    \hline   \hline
& \bf{MODEL DEPENDENT}    \\
\hline	    
		SMICA & Theory ($\Lambda$CDM ) masked  &  5.2 / 3 (15\%) & 60.2 / 3   ($<10^{-5}$)\\
	    $\Lambda$CDM Theory & Theory ($\Lambda$CDM ) masked   & 32.8 / 3 ($<10^{-5}$) & 0 (100\%)\\		    
	    $\Lambda$CDM  Sim. \#1 & Theory ($\Lambda$CDM ) masked  & 40.8 / 3 ($<10^{-5}$)  & 8.1 / 3  (4\%) \\
	    $\Lambda$CDM  Sim. \#2 & Theory ($\Lambda$CDM ) masked  & 56.7 / 3 ($<10^{-5}$)  & 5.4 / 3  (14\%) \\
	    $\Lambda$CDM  Sim. \#3 & Theory ($\Lambda$CDM ) masked  & 30.8 / 3 ($<10^{-5}$)  & 0.4 / 3  (94\%) \\
		\hline
	\end{tabular}
\end{table}

\begin{figure*}
     \begin{subfigure}[b]{1.\textwidth}
         \centering
         \includegraphics[width=\textwidth]{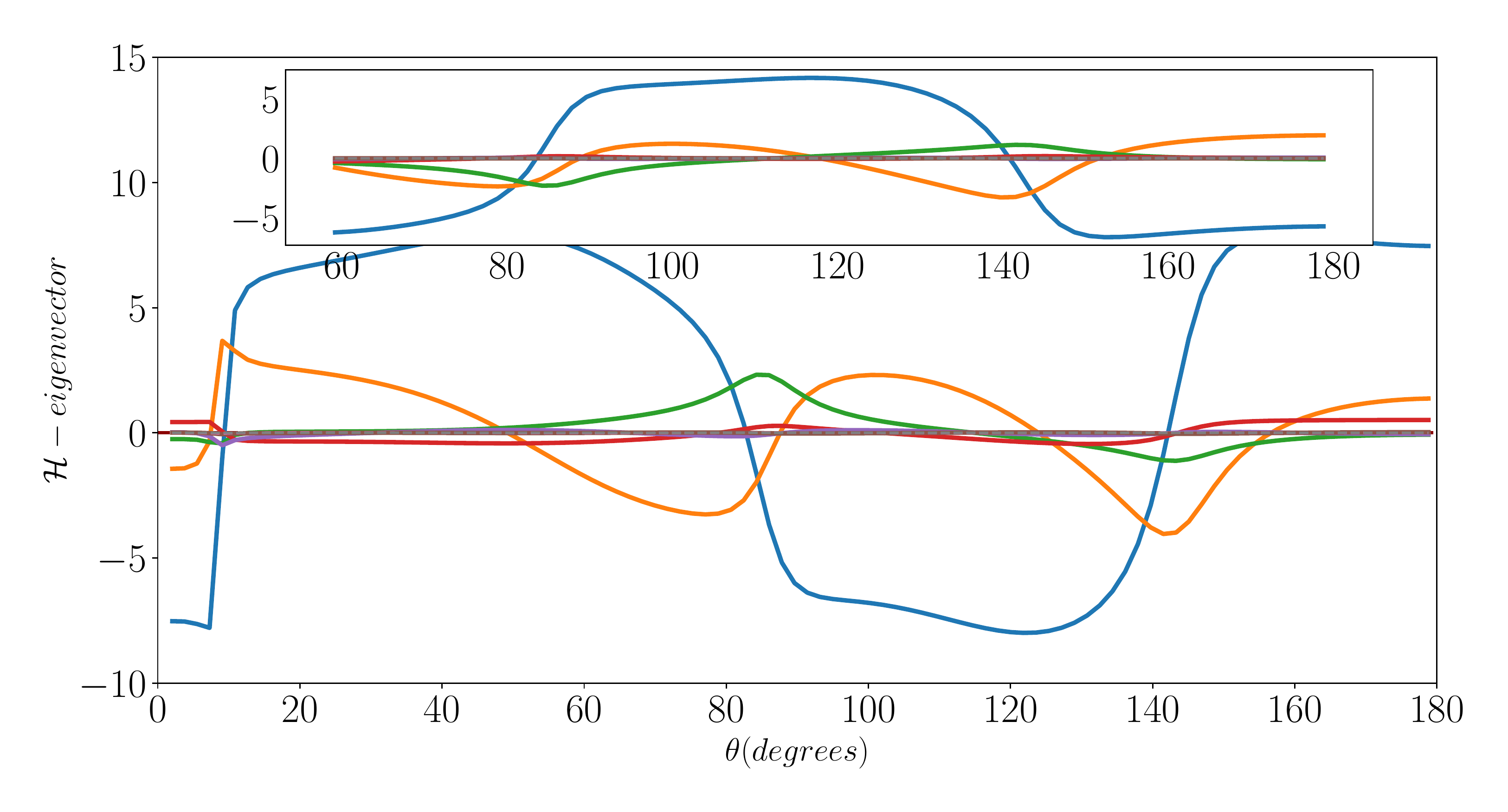}
         \caption{Weighted eigen-vectors.}
         \label{fig:H-eigen2}
     \end{subfigure}
     \begin{subfigure}[b]{1.\textwidth}
        \centering
         \includegraphics[width=\textwidth]{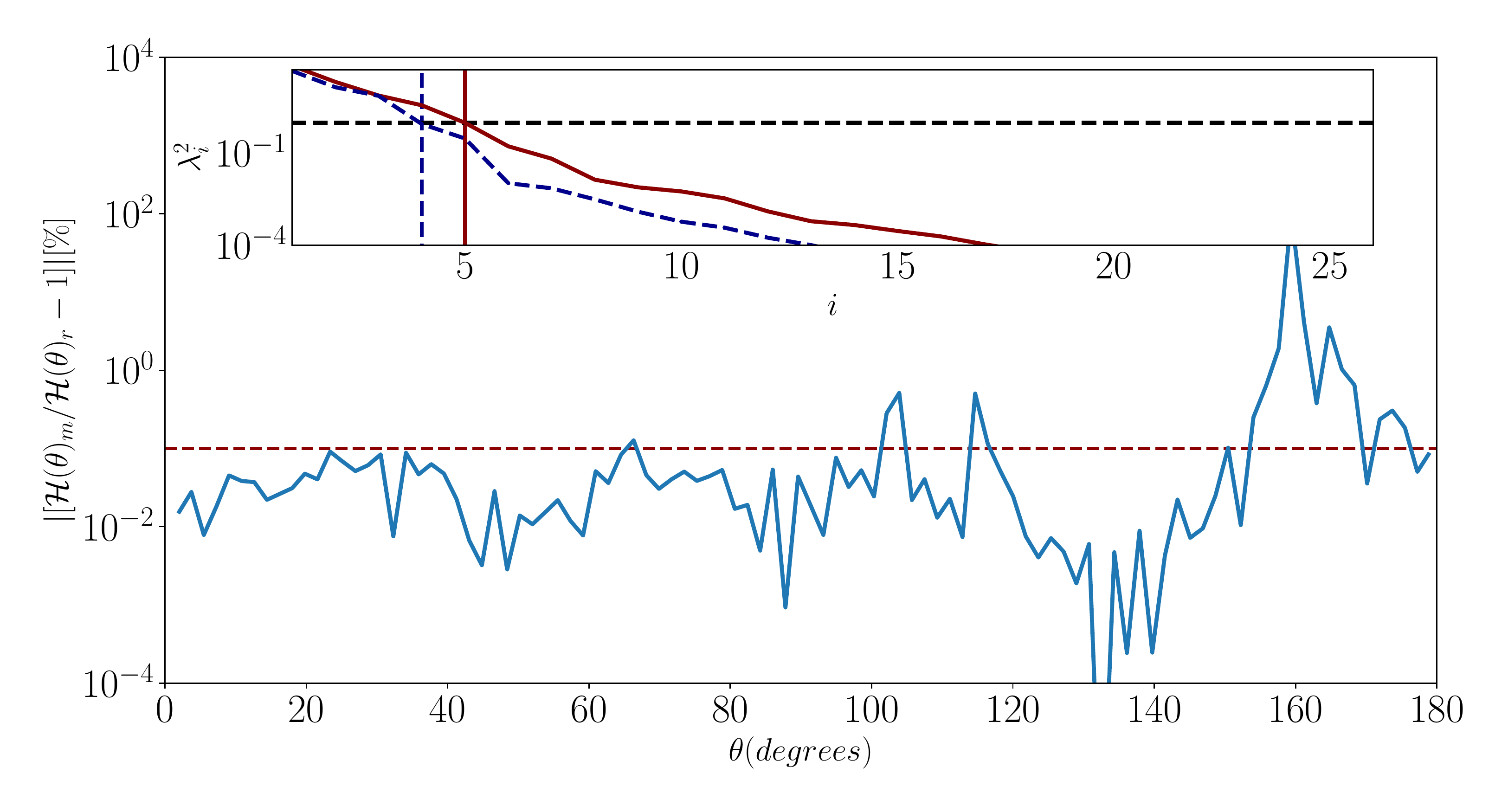}
         \caption{Comparison of the reconstructed and measured $\mathcal{H}(\theta)$.}
         \label{fig:H-rec-theory}
     \end{subfigure}
\caption{Principal Components Analysis (PCA) using Singular Value Decomposition (SVD) on the theoretical correlation matrix for $\mathcal{H}(\theta)$. This figure shows the same information that Fig.\ref{fig:PCA-results} but for the theoretical correlation matrix for $\mathcal{H}$.}
\label{fig:PCA-results-theo}
\end{figure*}

To test if the results are robust to different $N_{JK}$, we repeated the  analysis with  $N_{JK}=128$ using the same number of SV as for  $N_{JK}=64$ (to recover similar total $S/N$). In Table\ref{tab:PCA} we present both results and a very good agreement between them can be seen.  \textcolor{black}{We also show the results when running our pipeline on different reduction maps from Planck, which also show to be consistent between them.}

\begin{figure*}
    \centering
     \begin{subfigure}{0.32\textwidth}
         \centering
         \includegraphics[width=\textwidth]{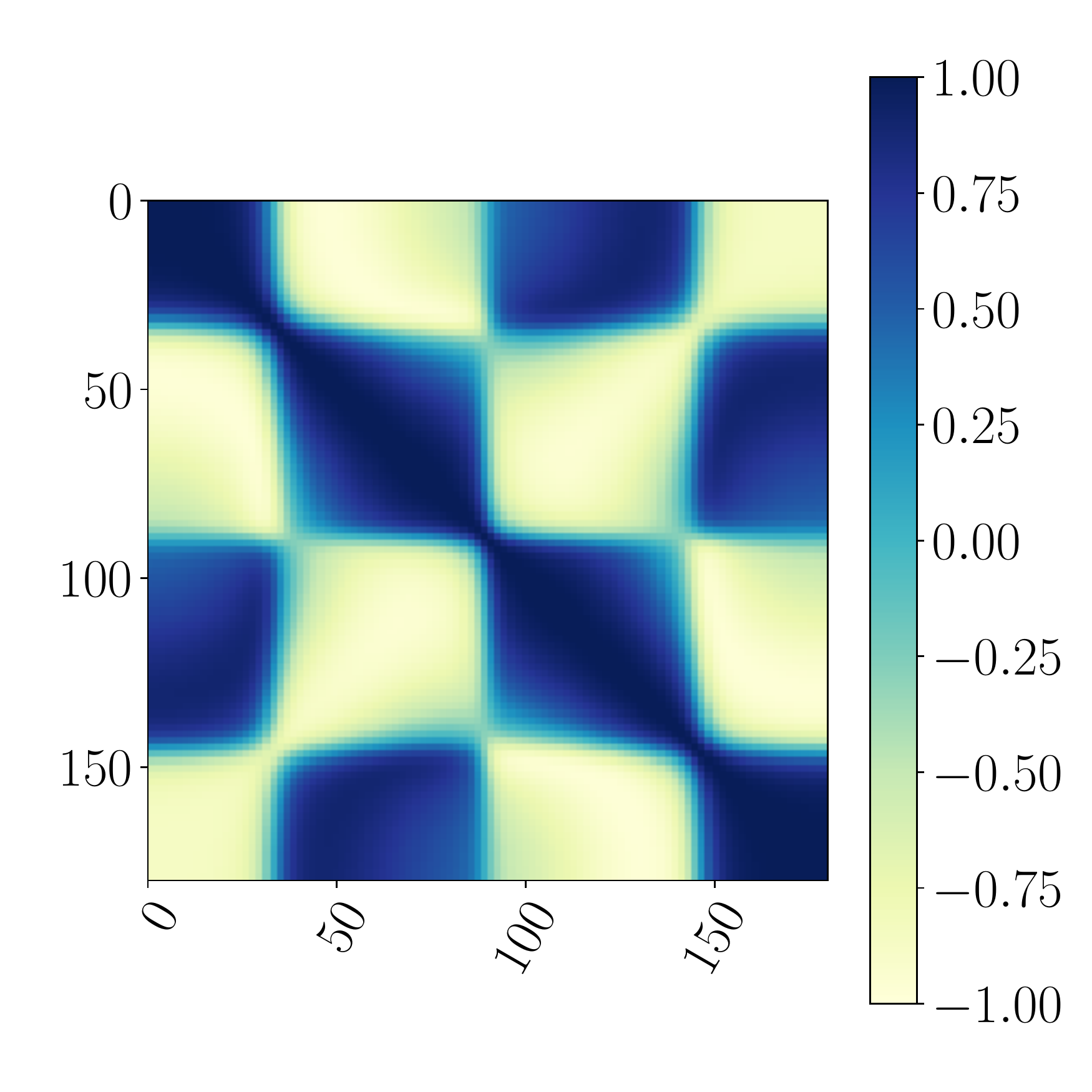}
         \caption{$\hat{C}_{ij}$ for $W_\eta(\theta)$.}
         \label{fig:w_theo_cov}
     \end{subfigure}
     \begin{subfigure}{0.32\textwidth}
        \centering
         \includegraphics[width=\textwidth]{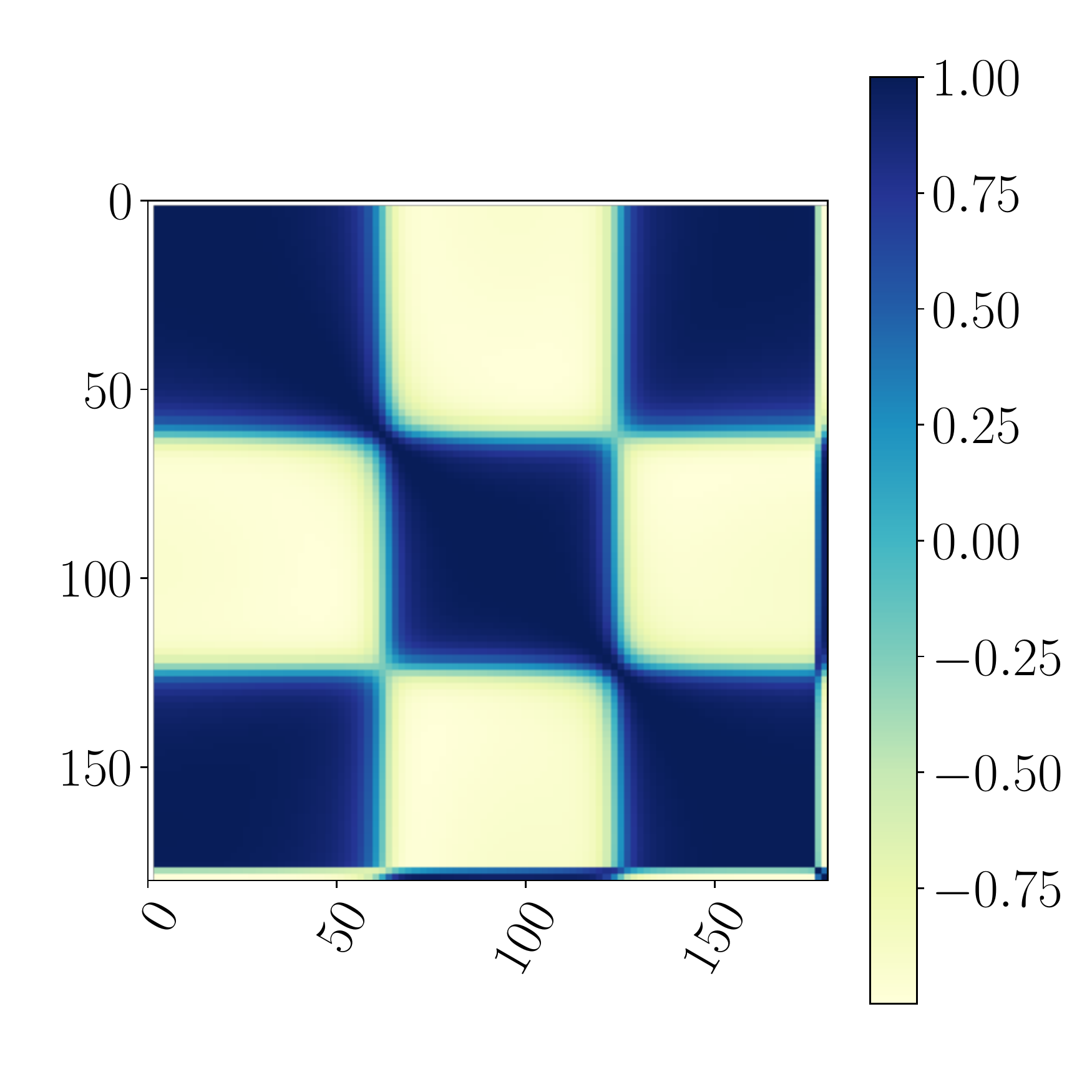}
         \caption{$\hat{C}_{ij}$ for $\mathcal{P}(\theta)$.}
         \label{fig:P_theo_cov}
     \end{subfigure}
     \begin{subfigure}{0.32\textwidth}
        \centering
         \includegraphics[width=\textwidth]{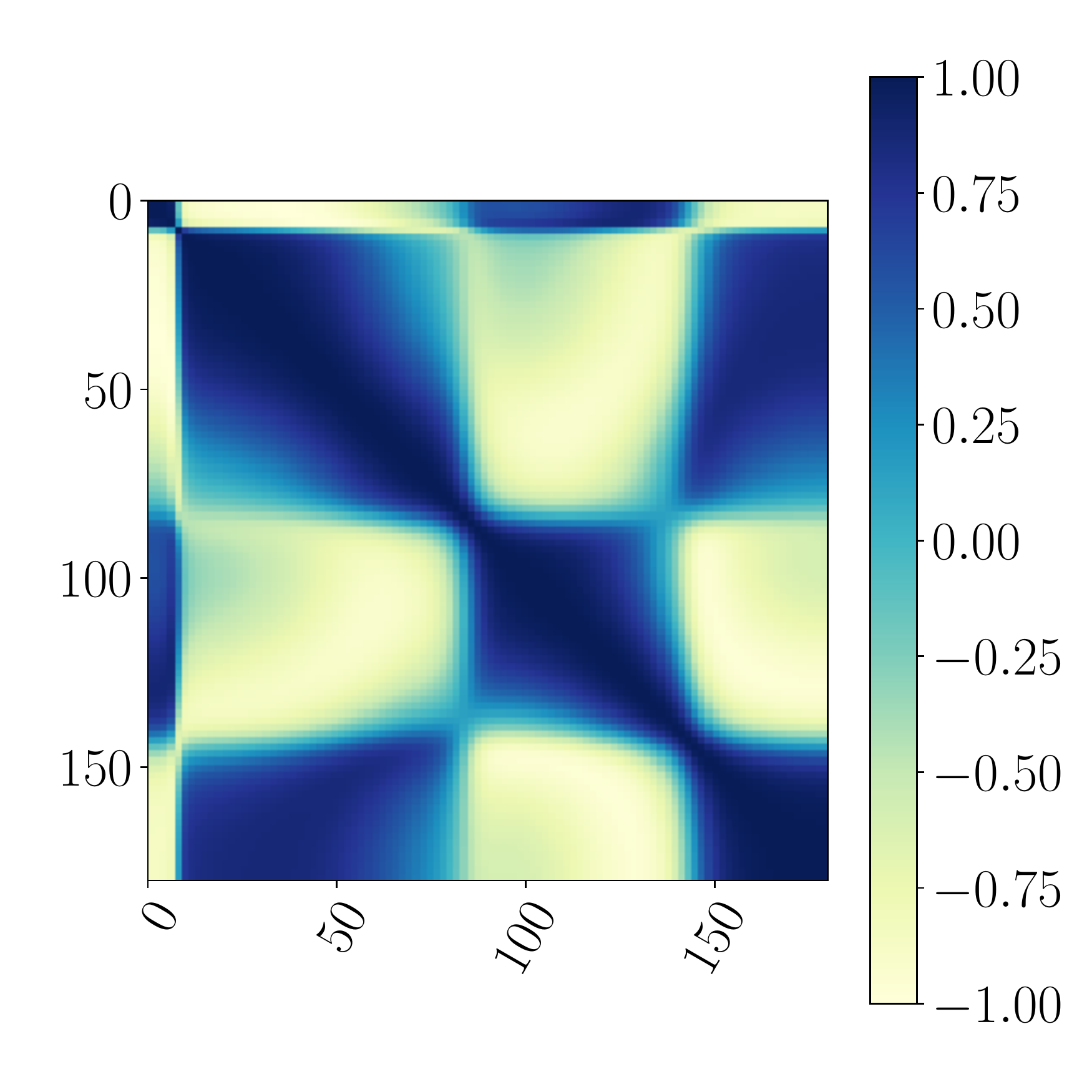}
         \caption{$\hat{C}_{ij}$ for $\mathcal{H}(\theta)$.}
         \label{fig:H_theo_cov}
     \end{subfigure}
    \caption{Theoretical Correlation matrices for $W_\eta(\theta)$, $\mathcal{P}(\theta)$ and $\mathcal{H}(\theta)$ estimated from expressions \ref{eq:theory-error}-\ref{eq:H_theory-cov}
    }
    \label{fig:theory-covariances}
\end{figure*}

\subsection{Analytic Error and its propagation}

Here we present an alternative estimate of the significance of our measurements to validate the JK approach. This is based on analytical estimates for the covariance matrices. This analytical expression has been shown to be a very good approximation for all-sky CMB measurements of $C_\ell$ in Eq.\ref{eq:transorm-cl2w}.
The  sampling variance covariance  of $C_\ell$
can be approximated as  being diagonal:
$Cov(\ell_i,\ell_j) \simeq
 \Delta^2 C_{\ell_i} \, {\cal{K}}_{ij}$, where ${\cal{K}}_{ij}$ is the Kronecker delta.
\cite{Cabre07} have shown, using simulations, that this is a good approximation with the variance:
\begin{equation}
    \Delta^2 C_\ell = \frac{1}{f_{sky}(2\ell+1)}\, C_\ell^2
\label{eq:theory-error}
\end{equation}
where $f_{sky}$ is the fraction of the sky which is not masked.
This allows us to compute the theoretical covariance in configuration space by just propagating errors from
Eq.\ref{eq:transorm-cl2w}:
\begin{equation}
C_{ij}^{W_\eta}(\theta_i,\theta_j) 
 = \sum_{\ell=1}\left( \frac{2\ell + 1}{4\pi}\right)^2
 P_{\ell}[\mu_i]
  P_{\ell}[\mu_j]
 \Delta^2C_{\ell}
    \label{eq:w_theory-cov}
\end{equation}
where $\mu_i \equiv \cos{\theta_i}$.
We can propagate the covariance in $W_\eta$ to that in $\cal{P}$ as follows:

\begin{equation}
C_{ij}^{\cal{P}}(\theta_i,\theta_j)  = \int_0^{\theta_i}  
 \frac{d\phi_i \sin{\phi_i}}{1-\cos{\theta_i}} 
 \int_0^{\theta_j} \frac{d\phi_j \sin{\phi_j}}{1-\cos{\theta_j}}  C_{ij}^{W_\eta}(\phi_i,\phi_j) 
\end{equation}
and from  $\cal{P}$ to $\mathcal H$:
\begin{equation}
C_{ij}^{\mathcal{H}} = 4 \frac{d^2
\tilde{C}_{ij}^{\cal{P}}
}{d\ln{\Omega_i} d\ln{\Omega_j}}
\label{eq:H_theory-cov}
\end{equation}
where $\tilde{C}_{ij}^{\cal{P}} \equiv C_{ij}^{\cal{P}}/({\cal{P}}_i {\cal{P}}_j)$. We can then estimate the corresponding correlation matrix: $\hat{C}_{ij} = C_{ij}/\sqrt{C_{ii} C_{jj}}$. 
Fig.\ref{fig:theory-covariances} shows the above analytical estimates using $C_\ell$ as measured by the Planck collaboration for the same CMB map we are using. 
Structures similar to the corresponding JK covariances in Fig.\ref{fig:covariances} can be seen, but with smaller number of independent blocks. This is probably related to the effects of the masks, which requires more  sub-structure to recover the same information.
Note that the theoretical covariance matrixes are also singular because of the strong correlation between angular bins. 
So we also perform a SVD to invert the covariance and estimate the $\chi^2$. This is shown in  Fig.\ref{fig:PCA-results-theo}. We keep the largest 4 SV for scales $\theta >59$ degrees, which corresponds to the same cut in the dominant values ($\lambda_i^ 2\gtrsim 0.2$, horizontal line in the sub-panel of the  Fig.\ref{fig:H-rec-theory}) as in the JK case.
The fact that there are a smaller number of SV for the theoretical covariance agrees with the visual appearance of the covariance matrix and the fact that masking in the real map requires more independent degrees of freedom to model the same signal. 

Table\ref{tab:PCA} compares the $\chi^2$ estimates. The significance is quite similar to JK covariance, indicating that our estimates are robust as to how we estimate the covariance and how we invert it. Also, we note how the total $(S/N)^2$ corresponds to the $\chi^2$ of the null model which in all cases is $> 100$ (for all angles). This is an indication that we are using the same equivalent number of SV modes in the different cases.

 \textcolor{black}{We note however that if we use the $\Lambda$CDM prediction as input $C_\ell$ (instead of the actual Planck measurements) to estimate the theoretical covariance, the  errors on $\mathcal{H}$  become much larger (because sampling variance errors are dominated by the amplitude of the quadrupole which is larger in  $\Lambda$CDM) and this results in only three independent SV and a lower $(S/N)^2$. In this reconstruction the
 $\Lambda$CDM model is still inconsistent with our measurements and with $\mathcal{H}=0$ for scales $>59$ degrees (last rows in the Table).}

 \subsection{Comparison with Simulations}
\label{sec:simulations}

\begin{figure*}
     \begin{subfigure}{0.5\textwidth}
         \centering
         \includegraphics[width=\textwidth]{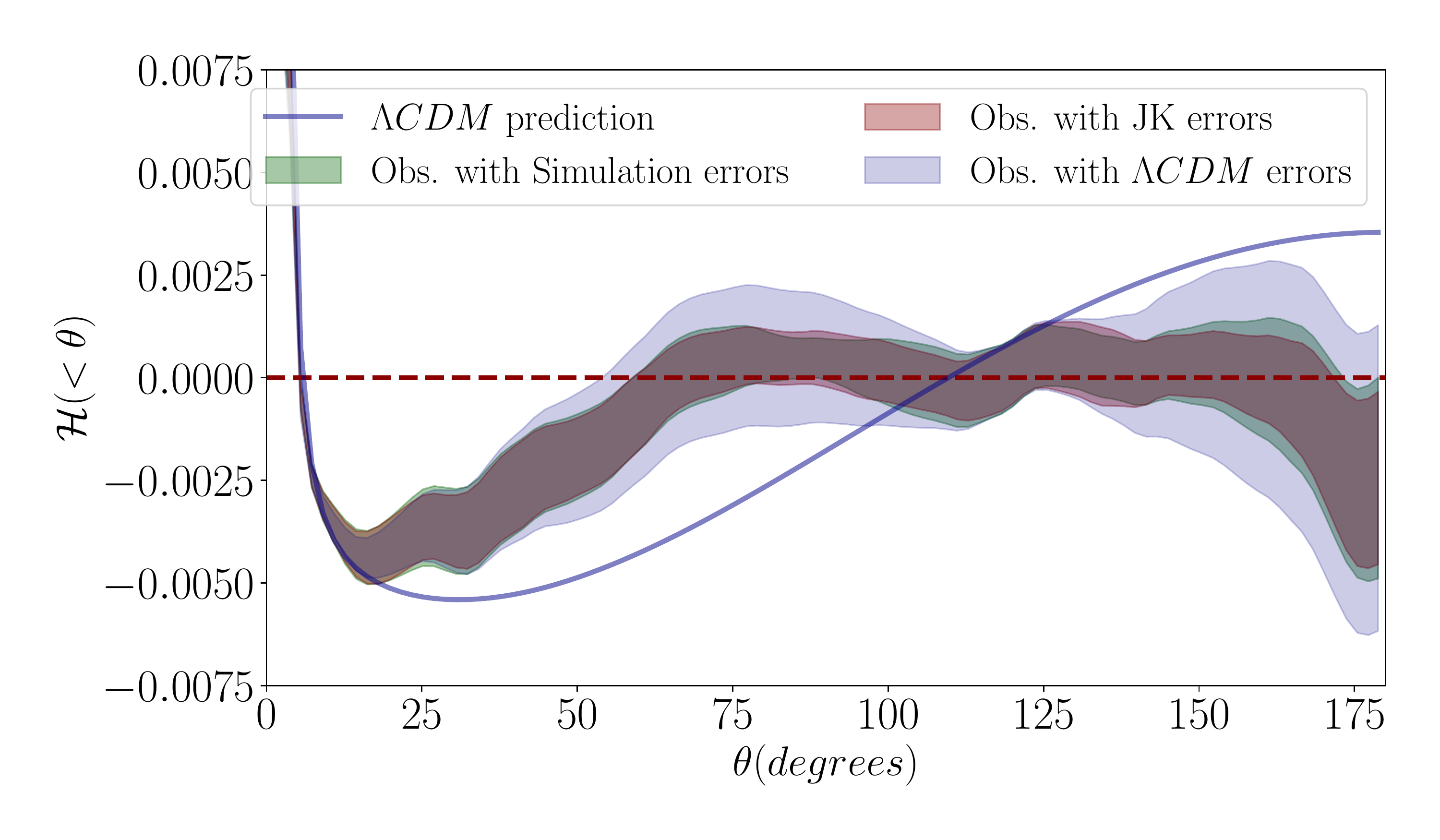}
         \caption{ $\mathcal{H}(\theta)$  comparison.}
         \label{fig:H-gaussian}
     \end{subfigure}
     \begin{subfigure}{0.5\textwidth}
        \centering
         \includegraphics[width=\textwidth]{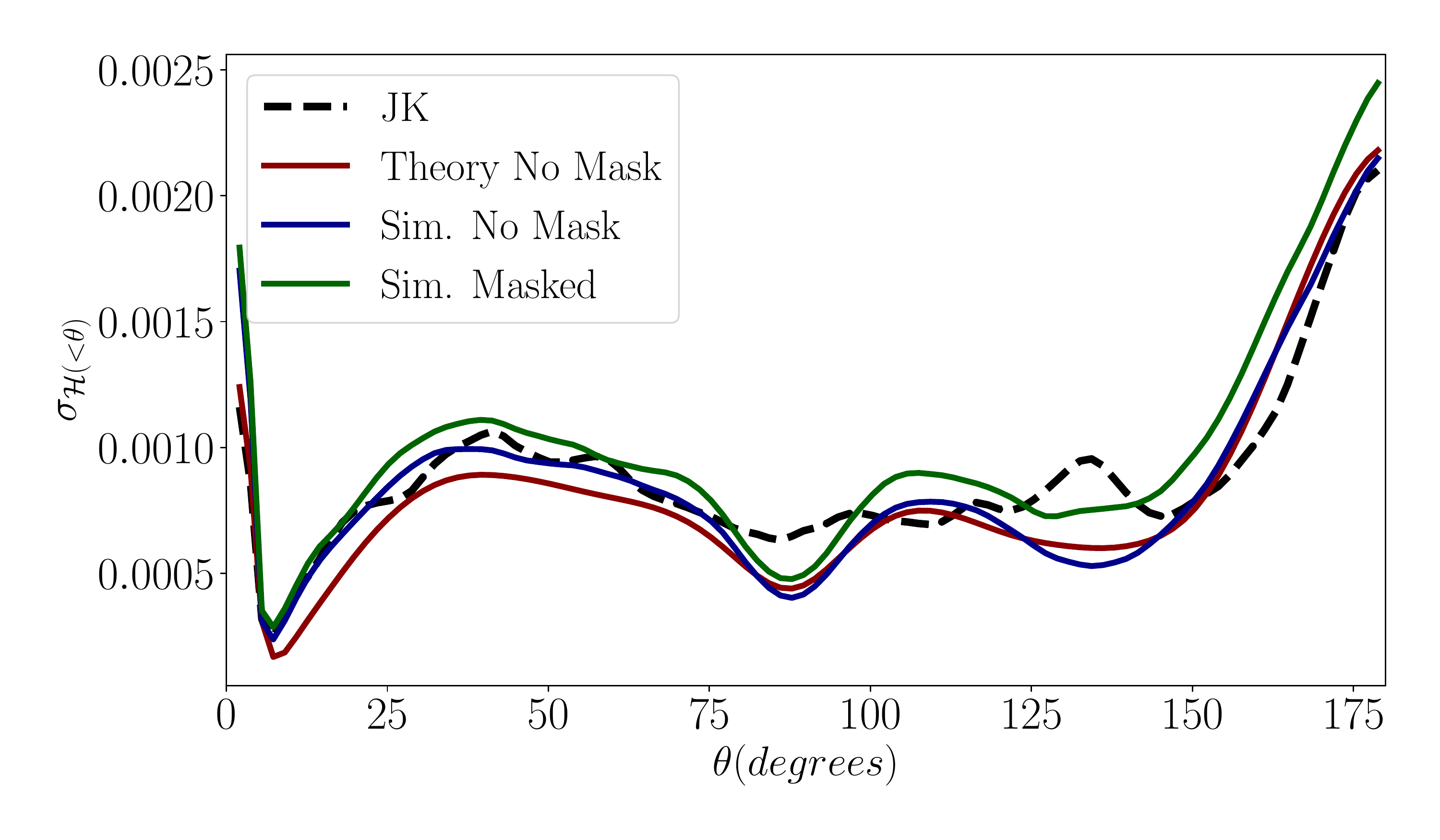}
         \caption{Diagonal error comparison $\sigma_{\mathcal{H}(<\theta)}$.}
         \label{fig:H-var-comparison}
     \end{subfigure}
\caption{$\mathcal{H}(\theta)$ and its standard deviation estimated from 128 simulations (Gaussian realizations) from the $C_\ell$s measured by the Planck collaboration. On the Fig.\ref{fig:H-gaussian}, the green solid line and green shaded area correspond to the mean and error from the simulations. The red shadow shows the mean and error-bars estimated directly from the CMB temperature maps with JK resampling. The solid black line, shows the theoretical prediction given by the Eq.\ref{eq:transorm-cl2w} and using the $C_\ell$ measured by Planck. The Fig.\ref{fig:H-var-comparison} shows the absolute error obtained by different pipelines: Jackknife, theoretical prediction and simulations. Where can be seen that there is a good agreement between all of them.}
\label{fig:H-Gaussian-realization}
\end{figure*}

 \textcolor{black}{
As a final validation of our pipeline we have done some simple simulations to test our measurements and error estimates. We have taken the measured angular power spectrum from Planck collaboration and made 128 realization implementing the \texttt{synfast} function from the \textit{healpy} python package. Then, we have computed the covariance between them. }

\begin{figure*}
    \centering
     \begin{subfigure}{0.32\textwidth}
         \centering
         \includegraphics[width=\textwidth]{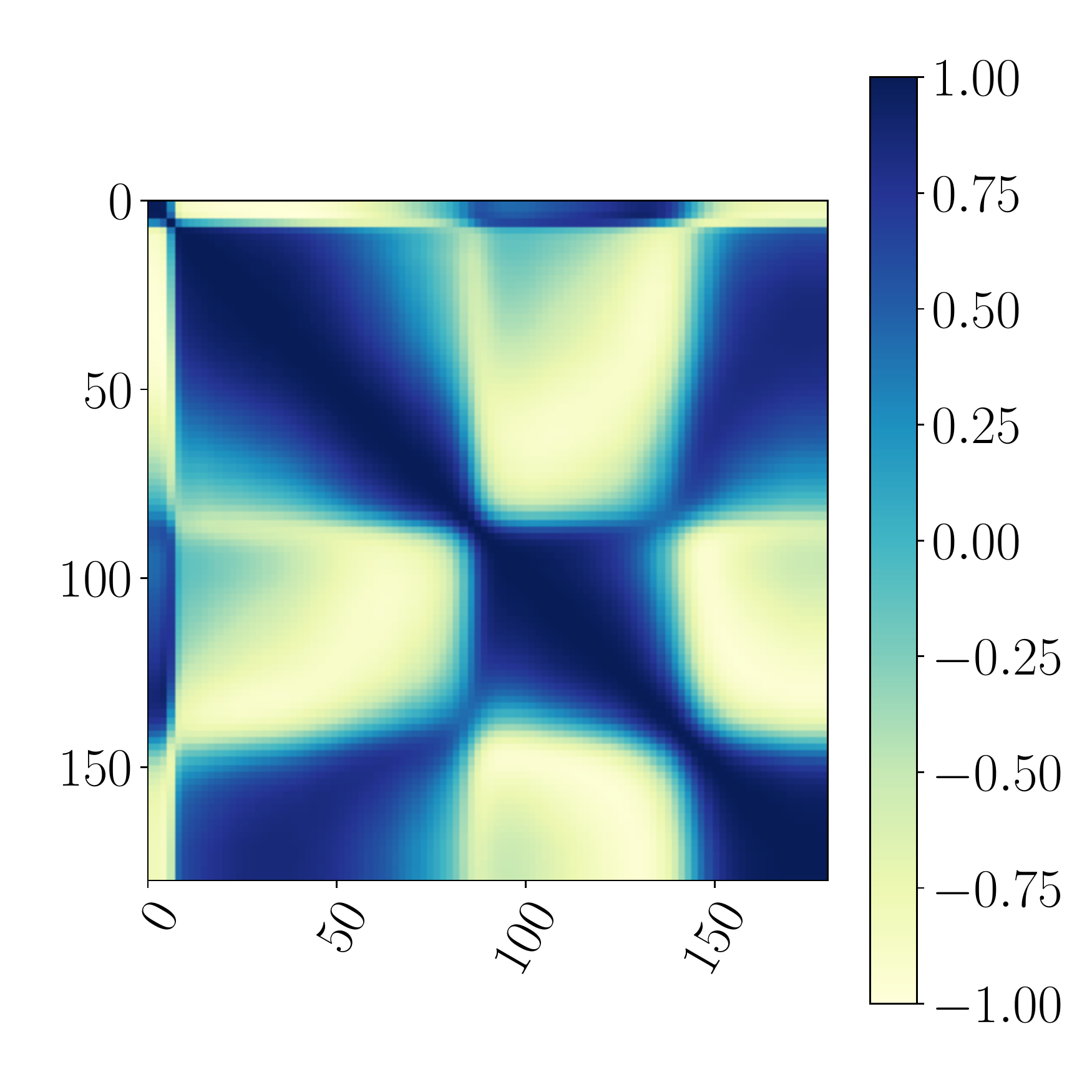}
         \caption{$\hat{C}_{ij}$ for $\mathcal{H}$ without mask.}
         \label{fig:H_gauss_cov}
     \end{subfigure}
     \begin{subfigure}{0.32\textwidth}
        \centering
         \includegraphics[width=\textwidth]{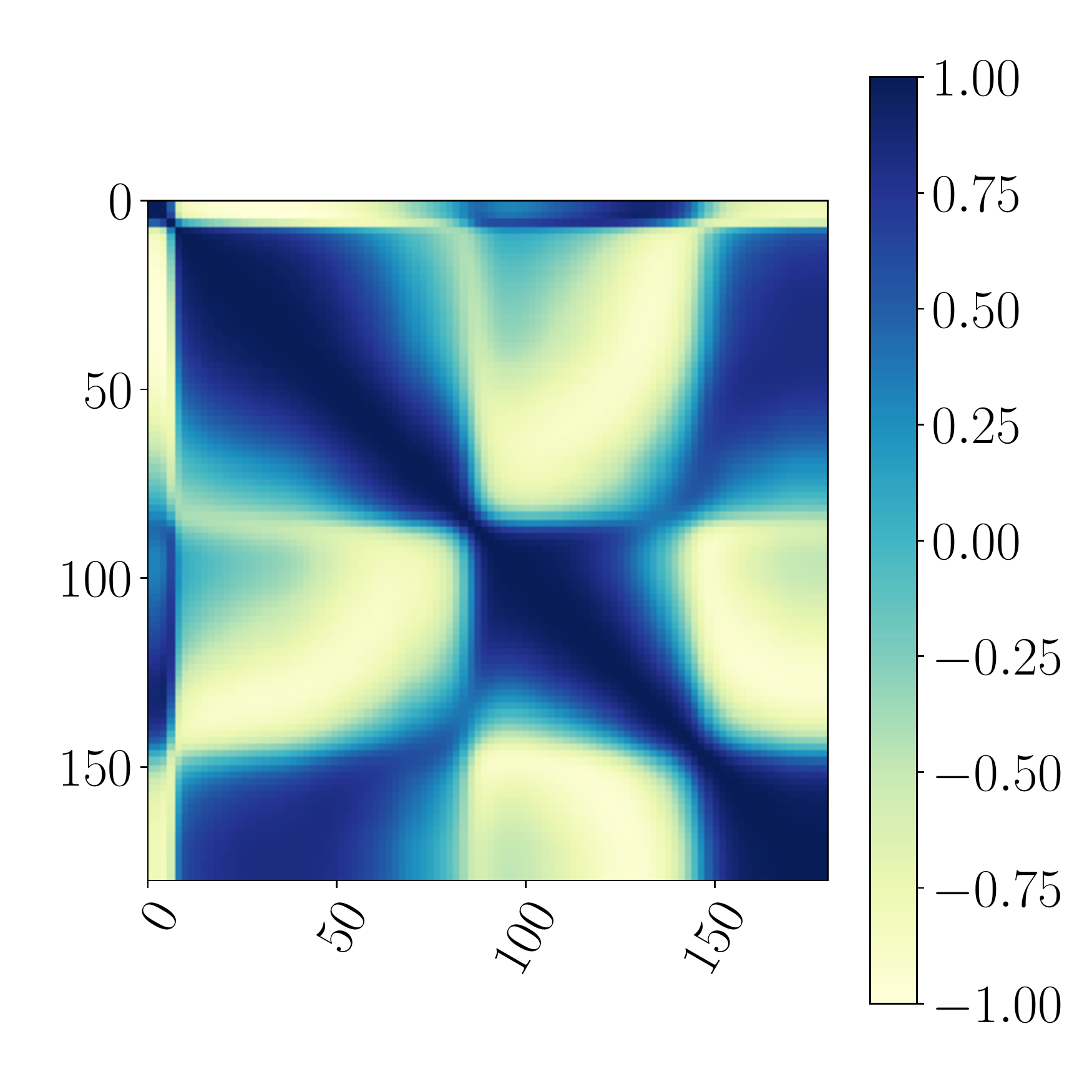}
         \caption{$\hat{C}_{ij}$ for $\mathcal{H}$ with mask.}
         \label{fig:H_gauss_cov_mask}
     \end{subfigure}
     \begin{subfigure}{0.32\textwidth}
        \centering
         \includegraphics[width=\textwidth]{images/H_theo_cov.pdf}
         \caption{$\hat{C}_{ij}$ for $\mathcal{H}$ from theory expressions.}
         \label{fig:H_theo_cov_2}
     \end{subfigure}
    \caption{Comparison of alternatives way of computing the correlation matrix for $\mathcal{H}(\theta)$. The Fig.\ref{fig:H_gauss_cov} show the correlation matrix of the simulations, which can be compared with the Figures \ref{fig:H_cov} and \ref{fig:H_theo_cov}.
    }
    \label{fig:theory-gaussian-covariances}
\end{figure*}

 \textcolor{black}{The resulting errors, as well as the correlation matrix, are show in Fig.\ref{fig:H-Gaussian-realization} and \ref{fig:theory-gaussian-covariances}. Panel \ref{fig:H-gaussian} compares measurement with the errors using different methods. There we can see that the amplitude of the errors obtained by JK or simulations are in good agreement.
The amplitude of the theoretical errors when computed with the $\Lambda$CDM model are larger, but the $\mathcal{H}$ predicted by the $\Lambda$CDM model (magenta shade) is still outside these larger errors. In order to test if this discrepancy is significant, we have computed the $\chi^2/d.o.f.$ using as errors the theoretical covariance obtained by the $\Lambda$CDM prediction. On Table \ref{tab:PCA}, we see that even implementing such larger errors, our measurement of  $\mathcal{H}$  is not compatible with the $\Lambda$CDM prediction. As expected, the same discrepancy is seen when comparing $\Lambda$CDM with the null model using the $\Lambda$CDM errors.}

 \textcolor{black}{On the Fig.\ref{fig:H-var-comparison} we show that errors (diagonal of the covariance) in $\mathcal{H}$ are very similar from JK, Theory or simulations with or without masks. Off diagonal elements (given by the correlation matrix) are also very similar as shown in \ref{fig:theory-gaussian-covariances}: the structure of the correlation matrix is similar to those obtained for the JK and theory cases on the panels \ref{fig:H_cov} and \ref{fig:H_theo_cov}.
This shows that the JK approach is adequate as previously demostrated by \cite{Cabre07}. It also shows that errors close to $180$ degrees increase by a factor of 3-4 which explains the large fluctuation in  $\mathcal{H}$  measured in the data at that scale. We can also see that the mask does not affect the results much.
 }
 
\begin{figure*}
         \centering
         \includegraphics[width=0.6\textwidth]{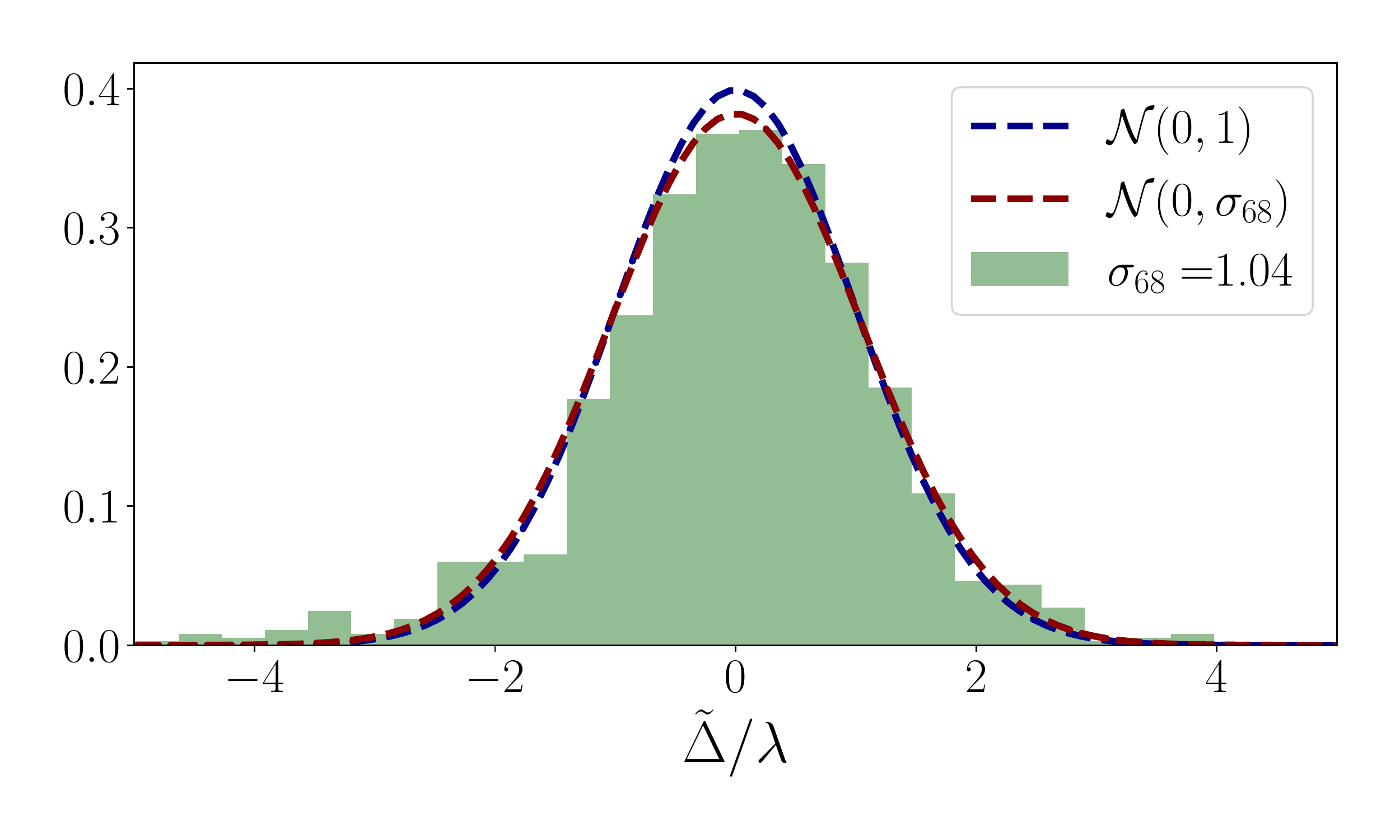}
         \caption{Distribution of $\hat{\Delta}_i/\lambda$ for the first 4 dominant components for the 256 Planck $C_\ell$ realizations. For comparison $\mathcal{N}(0,1)$ and $\mathcal{N}(0,\sigma_{68})$ are shown in the blue and red dasshed lines. } 
        
    \label{fig:D-Histogram}
\end{figure*}

 \textcolor{black}{ 
A complementary test allowed by doing simulations is to check if the distribution of the $\hat{\Delta}_i/\lambda$ values is Gaussian, in such a way that the $\chi^2$ can be easily interpreted. Fig.\ref{fig:D-Histogram} shows the distribution of $\hat{\Delta}_i/\lambda$ for the 4 dominant eigen-modes for 256 realizations with the $C_\ell$ measured by Planck. The blue line is a normal Gaussian distribution $\mathcal{N}(0,1)$, and the red line show $\mathcal{N}(0,\sigma_{68})$, where the $\sigma_{68}$ has been estimated from the histogram by doing a sigma clipping at the 16 and 84 percentiles. This exercise was done taking the scales $>59$, and illustrates well how such big scales are gaussianly distributed. }

 \textcolor{black}{ 
We can therefore interpret the $\chi^2$ values as probabilities $P$, shown in Table\ref{tab:PCA}. We find $P<10^{-5}$ for the $\Lambda$CDM model to be consistent with data ( $\mathcal{H}= \mathcal{H}_{\Lambda CDM}$) above 59 deg, when we use model independent covariances. This is also the case when  we use the covariance from the $\Lambda$CDM model. As a final test, in the last rows  we also test the $\Lambda$CDM prediction (model and simulations) with the $\Lambda$CDM covariance and find the expected results: i.e. that the  $\mathcal{H}$ values agree with  $\mathcal{H}= \mathcal{H}_{\Lambda CDM}$, but not with  $\mathcal{H}=0$ (contrary to the actual Planck data).} 

\subsection{Comparison with horizon predictions}
\label{sec:comparison}

\begin{figure}
    \centering
    \includegraphics[width=15cm]{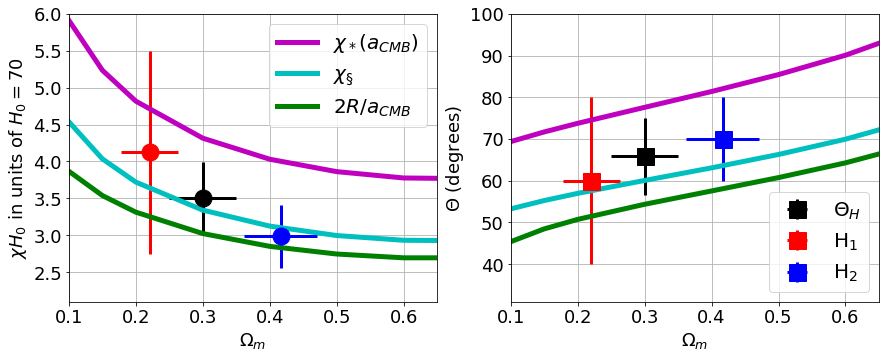}
    \caption{Comparison of the homogeneity scale  $\theta_{\mathcal{H}}$ from the homogeneity index (black dots) and  the causal horizon $H_1$ and $H_2$ in the CMB cosmological parameters \cite{Fosalba_2021},
    given the mean  measured $\Omega_m$ and $H_0$ in each region.
    Left panel shows comoving transverse radius  $\chi H_0$ (in units of $H_0=70$). Right panel shows the corresponding angular scale (independent of $H_0$).
     This is compared to the BHU predictions: $2R/a$ (green), $\chi_\S$
     \cite{Gazta_aga_2021} (cyan) and $\chi_*$ (magenta), at $a=a_{CMB}$ as a function of $\Omega_m$.}
    \label{fig:AnglePred}
\end{figure}

A given comoving scale $\chi$ can be observed in the CMB maps at an angle:
\begin{equation}
\theta_{CMB} = \frac{\chi}{d_A} 
\label{eq:angleCMB}
\end{equation}
where $d_A$  is the comoving radial lookback time coordinate to the CMB or the so call comoving  angular diameter distance to the CMB in Eq.\ref{eq:dA}.
Observations show that the expansion rate $H(\tau)$ today is dominated by $\rho_\Lambda \equiv \Omega_\Lambda \rho_c $, where $\rho_c= 3H_0^2/8\pi G$ is the critical density today. This means that the FLRW metric lives inside a trapped surface which is given by $r_\Lambda \equiv 1/H_\Lambda= (8\pi G\rho_\Lambda/3)^{-1/2} = \Omega_\Lambda^{-1/2}$ in units of $c/H_0$, which behaves like the interior of a BH \cite{gaztanaga:hal-03344159,gazta:Universe22}. 
To see this, consider outgoing radial null geodesic (the Event Horizon at $\tau$, \cite{Ellis1993,gaztanaga:hal-03344159}):
\begin{equation}
r_* \equiv a \chi_* = a(\tau) \int_\tau^\infty \frac{d\tau}{a(\tau)} =
a \int_a^\infty \frac{d\ln{a}}{a H(a)}  < \frac{1}{H_\Lambda} \equiv r_{\Lambda}
\label{eq:chi}
\end{equation}
where $\chi_*(a)$ is the corresponding comoving scale. For small $a$ the value of
$\chi_*$ is fixed to a constant $\chi_* \simeq 3r_\Lambda$. Thus, the physical trapped surface radius $r_*$ increases with time. As we approach $a \simeq 1$ the Hubble rate becomes constant and $r_*$ freezes to a constant value $r_* = r_\Lambda$
(see  Fig.2 in \cite{gaztanaga:hal-03344159}).

Fig.\ref{fig:AnglePred} shows the prediction for $\theta_{CMB}$ for $\chi_*$ in the CMB as a function of $\Omega_m$, just as a reference.
Our result for $\theta_{CMB}$ 
is in agreement  with 
Fig.31 in \cite{Fosalba_2021} which shows  the
size of CMB horizons in measured variations of cosmological parameters. Each causal horizon of similar cosmological parameters  has a different size  $\theta_{CMB}$ and different mean cosmological paramters (e.g. $H_0$ and $\Omega_m$). In Fig.\ref{fig:AnglePred} we show regions labeled $H_1$ and $H_2$ in \cite{Fosalba_2021} which are the two largest and more significant.

We compare these measurements with the prediction for the causal comoving horizon $\chi_\S$:

\begin{equation}
\rho_{\Lambda} =  <\rho_m/2+\rho_R>_{\chi_\S}, 
\label{eq:chiC}
\end{equation}
which is defined as the comoving radial coordinate that gives zero action boundary $S^{on-sh}=0$ for the evolution of our Universe \cite{Gazta_aga_2021}. This causal scale $\chi_\S$ is slightly smaller than the above trapped surface $\chi_*$. We also compared the results with the comoving 
cut-off $2R/a_{CMB}$ in super-horizon perturbations predicted by the BHU model \cite{gaztanaga:hal-03344159,gazta:Universe22} which corresponds to the radius $R=[r_\Lambda r_H^2]^{1/3}$ of the FLRW cloud which collapsed and bounced into the Big Bang expansion.
Fig.\ref{fig:AnglePred} shows that the two predictions
$2R/a_{CMB}$ and $\chi_\S$ scales are consistent with our estimate. The later seems preferred but the errors are too large to distinguish well among these models at 2 or 3 sigma significance level.

Note how $\Omega_m \simeq 0.3$ is preferred by our $\theta_{CMB}$ homogeneity index measurements in combination with the horizon predictions. 
Such value is similar to measurements by many other cosmological observations, such as SNIa, BAO and ISW (e.g. see \cite{Gaztanaga2006,Gaztanaga2009,DiValentino,des2018} and references therein), but is totally independent from them. This indicates that our measurement of the homogeneity scale  $\theta_{\mathcal{H}}$ can indeed be interpreted as a cut-off in super horizon perturbations and therefore as a sign of homogeneity. The $\Lambda$CDM model has more power on larger scales because super horizon perturbations from Inflation don't have such a cut-off and the Universe is assumed to be much larger and more irregular.


\section{Discussion and Conclusions}
\label{sec:Conclusions}


We have estimated the scale of homogeneity in the Planck CMB temperature maps using the homogeneity index (equivalent to the Haussdorf dimension) in Eq.\ref{Eq:homogeneity-index-T}.
We find that such early universe starts to be homogeneous at scales larger than 60 degrees, which is not in agreement with the standard $\Lambda$CDM prediction.

Fig.\ref{fig:H-linear} shows how, contrary to real data,  $\mathcal{H}$ for 
the  $\Lambda$CDM  model does not tend to zero.  This is because geometrical homogeneity is never reached for  $\Lambda$CDM because
it assumes scale invariance (i.e. a fractal structure) over all scales. This indicates that the nature (and origin) of such perturbations could be different than assumed.
This new result is not just a rephrasing of previous estimates based on $w(\theta)$
(e.g. see \cite{Copi2009,Efstathiou2010} and references therein). It uses a geometrical measurement, $\mathcal{H}$ , which contrary to $w(\theta)$ is independent of the amplitude of CMB temperature fluctuations. For $W_\eta=1+w(\theta)$
we find that correlation nulls around scales of $\theta_{CMB} \simeq 70 \pm 16$ degrees, consistent with previous estimates. For the (fractal or Hausdorff) amplitude independent homogeneity index $\mathcal{H}=0$ we find instead:
$\theta_{CMB} \simeq  66 \pm 9$, which has smaller error-bars.

We evaluate the significance of our measurements  using a PCA analysis in a model-independent way, i.e. using errors based on data. The PCA is needed because we have to use many angular bins to be able to have a good estimate of the homogeneity index and those bins are strongly correlated. Therefore the corresponding covariance is singular as there are many degenerate degrees of freedom. We find  strong evidence for a Universe that is homogeneous above 60-70 degrees. 

 \textcolor{black}{When we use errors based on data, we find that the
$\Lambda$CDM best fit model is about  $3\sigma$ away (per degree of freedom) from the  homogeneity index measured directly on the temperature map (Table \ref{tab:PCA}). 
The $\Lambda$CDM predictions have more power on the low $C_L$ power spectrum multipoles $L$ than the actual CMB measurements. This results in very large sampling variance error predictions for $w(\theta)$ and also for  $\mathcal{H}$. Nevertheless, the $\Lambda$CDM inconsistency with the measured $\mathcal{H}$ is also significant if we use model dependent errors base on the  $\Lambda$CDM prediction  (bottom entries in Table \ref{tab:PCA}).}

CMB maps also include other contaminants  such as the  ISW,  Sunyaev-Zeldovich and lensing contributions (\cite{FGC2003,ISW}). Lensing and ISW effect have a very broad radial kernel so that anisotropies at a fixed redshift are washed away. These have therefore been ignored in our interpretation, but further work is needed to assess their contribution. 

These results are in agreement with a
recent study made by \cite{Fosalba_2021}, where they found evidence for causal horizons on scales around $\simeq 60-80$ deg. Note how these results are quite independent. In our case we are measuring the scaling of normalized temperature fluctuations on angular scales larger than 60 degrees, whereas \cite{Fosalba_2021}  find spatial variations of the cosmological parameters fitted to the BAO physics on degree and smaller scales (averaged over on large regions of the sky).

 \textcolor{black}{Cosmic expansion is dominated by $\rho_\Lambda$ and this indicates that we live inside a trapped surface (or event horizon) given by Eq.\ref{eq:chi}-\ref{eq:chiC}.  Our universe behaves like the inside of a BH \cite{Gazta_aga_2021, gaztanaga:hal-03344159,gazta:Universe22} where the measured $\rho_\Lambda$ corresponds to the BH event horizon \cite{Gazta2023sym}.
The results we find here, fit better within this bounded Black Hole Universe (BHU) cosmology \cite{gaztanaga:hal-03344159,gazta:Universe22} than with the $\Lambda$CDM prediction (from Inflation), which is scale invariant and has no cut-off scale, i.e. it is not homogeneous at any scale.
For $\Omega_\Lambda \simeq 0.7$ the trapped surface correspond to  $60-70$ degrees on the CMB sky, as shown in Fig.\ref{fig:AnglePred}  (see also \cite{Gazta_aga_2020,Gazta_aga_2021,gazta:Universe22,gaztanaga:hal-03344159,GF21}).
This shows that the origin of the large scale homogeneity  $\mathcal{H}$ could be the event horizon of the BHU model.}

\acknowledgments

 This work has been supported by spanish MINECO  grants PGC2018-102021-B-100 and EU grants LACEGAL 734374 and EWC 776247 with ERDF funds. BC 
 acknowledge support from a PhD scholarship from the Secretaria d’Universitats i Recerca de la Generalitat de Catalunya i del Fons Social Europeu.
IEEC is funded by the CERCA program of the Generalitat de Catalunya. This work has been carried out within the framework of the doctoral program of Physics of the Universitat Autònoma de Barcelona.




\clearpage
\printbibliography

\end{document}